\documentclass[preprint]{aastex631}
\usepackage[normalem]{ulem}
\usepackage{textcomp, gensymb, amsmath}
\useunder{\uline}{\ul}{}

\accepted{March 23, 2024}

\submitjournal{ApJ}

\begin{document}

\title{Short-term Classification of Strong Solar Energetic Particle Events using Multivariate Time Series Classifiers}

\correspondingauthor{Sumanth A. Rotti}
\email{srotti@gsu.edu}

\author[0000-0003-1080-3424]{Sumanth A. Rotti}
\affiliation{Georgia State University \\
Department of Physics and Astronomy \\
Atlanta, GA, USA}

\author[0000-0002-9799-9265]{Berkay Aydin}
\affiliation{Georgia State University \\
Department of Computer Science \\
Atlanta, GA, USA}

\author[0000-0001-8078-6856]{Petrus C. Martens}
\affiliation{Georgia State University \\
Department of Physics and Astronomy \\
Atlanta, GA, USA}

\begin{abstract}
Solar energetic particle (SEP) events are one of the most crucial aspects of space weather that require continuous monitoring and forecasting. Their prediction depends on various factors including source eruptions. In the present work, we use the Geostationary Solar Energetic Particle (GSEP) data set covering solar cycles 22, 23, and 24. We develop a framework using time series-based machine learning (ML) models with the aim of developing robust short-term forecasts by classifying SEP events. For this purpose, we introduce an ensemble learning approach that merges the results from univariate time series of three proton channels (E$\geq$10 MeV, 50 MeV, and 100 MeV) and the long band X-ray flux (1–8Å) channel from the Geostationary Operational Environmental Satellite (GOES) missions and analyze their performance. We consider three models, namely, time series forest (TSF), supervised time series forest (STSF) and bag of SFA symbols (BOSS). Our study also focuses on understanding and developing confidence in the predictive capabilities of our models. Therefore, we utilize multiple evaluation techniques and metrics. Based on that, we find STSF to perform well in all scenarios. The summary of metrics for the STSF model is as follows: AUC = \textbf{0.981}; F$_{1}$-score = \textbf{0.960}; TSS = \textbf{0.919}; HSS = \textbf{0.920}; GSS =  \textbf{0.852}; and MCC = \textbf{0.920}. The Brier score loss of the STSF model is \textbf{0.077}. This work lays the foundation for building near-real-time (NRT) short-term SEP event predictions using robust ML methods.
\end{abstract}


\section{Introduction} \label{sec:intro}
Solar energetic particle (SEP) events are manifestations of solar activity that constitute the emission of energetic electrons, protons and heavier ions from the Sun. These events are usually associated with parent solar eruptions, namely solar flares (SFs) and shock fronts of coronal mass ejections (CMEs; Cane et al. \citeyear{1986cane}; Kahler \citeyear{1992kahler}; Reames \citeyear{reames1999particle}; Gopalswamy et al. \citeyear{gopalswamy2001predicting}). Generally, it is understood that the eruptions at the western side of the Sun have a higher probability of SEPs reaching near-Earth space due to the spiral structure of the interplanetary magnetic field lines, known as the Parker spiral (Parker \citeyear{parker1965dynamical}; Reames \citeyear{reames1999particle}). Measurements of SEP events near Earth depend on the spatial region of source eruptions on the Sun. In the case of extreme SEP events, given the right conditions such as geomagnetic connectivity and enough seed population, they are often associated with fast CMEs (Marqu{\'e} et al. \citeyear{marque2006solar}; Gopalswamy et al. \citeyear{gopalswamy2008radio}; Swalwell et al. \citeyear{swalwell2017solar}; Gopalswamy et al. \citeyear{gopalswamy2017hierarchical}; Cliver \& D'Huys \citeyear{cliver2018size}; Rotti \& Martens \citeyear{rotti2023analysis}). 

The impacts of SEP events include severe technological (Smart \& Shea \citeyear{smart1992}) and biological effects on various economic scales (Schrijver \& Siscoe \citeyear{schrijver2010heliophysics}). Although the Earth's magnetic field provides us a protective shield from the energetic particles and filters them out from reaching the ground, they can be fatal for space-based missions and aircraft travel along polar routes (Beck et al. \citeyear{beck2005tepc}; Schwadron et al. \citeyear{schwadron2010earth}). For instance, long-lasting strong SEP events pose a radiation hazard to astronauts and electronic equipment in space (Jiggens et al. \citeyear{jiggens2019situ}).

According to the Space Weather Prediction Center (SWPC) proton intensities $\ge$10 pfu (1 pfu = 1 particle per cm$^2$.s.sr) in the E$>$10 Mega electron-Volt (MeV) energy channel are termed as large SEP events with regards to causing significant space weather (SWx) effects (Bain et al. \citeyear{bain2021summary}). In addition, the severity of the solar proton events is measured by SWPC using the Solar Radiation Storm Scale (S-scale)\footnote{\url{https://www.swpc.noaa.gov/noaa-scales-explanation}} which relates to biological impacts and effects on technological systems. The S-scale relies on the E$\ge$10 MeV integral peak proton flux from
near-Earth observations of the Geostationary Operational Environmental Satellite (GOES) missions (Sauer \citeyear{sauer1989sel}; Bornmann et al. \citeyear{bornman}). The base threshold, associated with an S1 storm, corresponds to a GOES five minutes averaged $\ge$10 MeV integral proton flux exceeding 10 pfu for at least three consecutive readings. Further scales from `S2' to `S5' logarithmically increase from one another, therefore defining different event intensities.

With great advancements in space engineering and technology, we are fortunate to have near-continuous observations of solar activity from a fleet of space-based satellites over the last four decades. One important aspect of analyzing solar data is to advance operational capabilities by mitigating SWx effects on our human explorers and technological systems (Jackman \& McPeters \citeyear{jackman1987solar}). This urgently requires the development of robust tools to forecast eruptive event occurrences. With an SEP event prediction system we can forecast and send out warning signals before the event.

Several researchers have been focusing on implementing a variety of model-driven techniques for predicting SEP events. In this regard most scientific studies concentrate on predicting the peak fluxes. To predict event occurrences, many physics-based and data-driven statistical models have been designed based on the parameters of parent eruptions such as SFs and CMEs (Van Hollebeke et al. \citeyear{1975van}; Posner \citeyear{posner2007up}; Kahler et al. \citeyear{2007kahler}; Balch \citeyear{balch}; Laurenza et al. \citeyear{laurenza2009technique}; {N{\'u}{\~n}ez} \citeyear{2011nunez}; Falconer et al. \citeyear{falconer2011tool}; Dierckxsens et al. \citeyear{dierckxsens2015relationship}; Winter \& Ledbetter \citeyear{winter2015ApJ}; {N{\'u}{\~n}ez} \citeyear{2015nunez}; Anastasiadis et al. \citeyear{anastasiadis2017predicting}; Alberti et al. \citeyear{alberti2017solar}; Papaioannou et al. \citeyear{papaioannou2018nowcasting}; Ji et al. \citeyear{Ji2021}). In the last decade machine learning (ML) methods have also been at the forefront of SEP event forecasting (Swalwell et al. \citeyear{swalwell2017solar}; Engell et al. \citeyear{engell2017sprints}; Aminalragia-Giamini et al. \citeyear{aminalragia2021solar}; Lavasa et al. \citeyear{lavasa2021assessing}). ML-based algorithms have been rigorously explored by many teams across the globe due to their success in many other areas of research and operations (Camporeale \citeyear{enrico2019}). Detailed descriptions of existing SEP event forecasting models can be found in Whitman et al. (\citeyear{WHITMAN2022}).

We envision building low-risk, short-term predictive models as the first step towards building operationally driven, reliable SEP event forecasting systems. Therefore, we exploit the feasibility of multivariate time series (MVTS) data in this work. For this purpose, we utilize and compare the performances of three ML models. Two are interval-based algorithms: time series forest (TSF) and supervised time series forest (STSF)—lastly, a dictionary-based bag of SFA symbols (BOSS) model. Prior studies on SEP event forecasting using parent eruption features conclude that the tree-based model is viable (Boubrahimi et al. \citeyear{boubrahimi2017prediction}). Both TSF and STSF implement a highly specialized random forest (RF) model and rely on several interpretable statistical features extracted from the time series to feed into an ensemble of decision trees. We will discuss more on individual model architectures in the later part. The rest of the paper is organized as follows: Section \ref{sec:data} provides information about our data set and data preparation steps used in this work. Section \ref{sec:methods} presents our research methodology including descriptions of the time series classifiers. Section \ref{sec:results} discusses the training phase of the models and presents the experimental evaluation framework. Lastly, Section \ref{sec:conclusion} summarizes our work and future avenues.

\section{Data} \label{sec:data}
The SEP events are critical phenomena caused by SFs and CMEs. The parent eruptions are triggered by sudden, abrupt changes in the magnetic field, typically of active regions in the solar atmosphere. Thus, it is well expected to build predictive capabilities employing parameters of precursor events. Nonetheless, we do not consider any data related to CMEs, and restrict ourselves to use the one-minute averaged GOES X-ray (1–8Å) fluxes measured by the X-ray sensor (XRS) onboard GOES. The archived data is available online from the National Oceanic and Atmospheric Administration (NOAA)'s website\footnote{\url{https://www.ncei.noaa.gov/data/goes-space-environment-monitor/access/avg/}}. In addition, we use the following integrated proton channels from GOES: (1) E$\geq$10 MeV fluxes corresponding to P3, (2) E$\geq$50 MeV fluxes corresponding to P5, and (3) E$\geq$100 MeV fluxes corresponding to P7. Because SFs have characteristic durations from a few minutes to a few tens of minutes, we linearly interpolate the proton five-minute averaged fluxes to match with the one-minute cadence of the X-ray fluxes. We believe this interpolation is necessary to retain the information on flaring peaks without altering the flare characteristics from X-ray fluxes.

\subsection{GSEP Data Set}
The Geostationary Solar Energetic Particle (GSEP) events data set (Rotti et al. \citeyear{gsep_2022}) is a recently introduced open-source\footnote{The GSEP data set available on Harvard Dataverse: \dataset[10.7910/DVN/DZYLHK]{https://doi.org/10.7910/DVN/DZYLHK}} multivariate time series (MVTS) benchmark data set of SEP events covering solar cycles 22 to 24. The description of the data set and its development can be found in Rotti et al. (\citeyear{rotti2022}) and Rotti \& Martens (\citeyear{rotti2023analysis}). It was created using proton fluxes measured by the Space Environment Monitor (SEM) suite onboard GOES (Grubb \citeyear{grubb}). This data set comprises a catalog of 433 (- 244 large and - 189 small) SEP events observed near Earth between 1986-2018. Each event is labeled a `1' or `0', indicating either a large or small SEP event, respectively. \deleted{based on the event definition of NOAA-SWPC.} Here, a large SEP event corresponds to proton fluxes crossing 10 pfu in the GOES `P3' channel. Whereas a small SEP event has proton enhancements between $\geq$0.5 and $<$10 pfu. Furthermore, the data set consists of time series slices of GOES proton and X-ray fluxes of all the events. Each time series slice constitutes 12 hr fluxes prior to the onset of the event as an observation window and further, until the events cross the peak flux, finally falling to half that value.

\begin{figure}[ht!]
\plotone{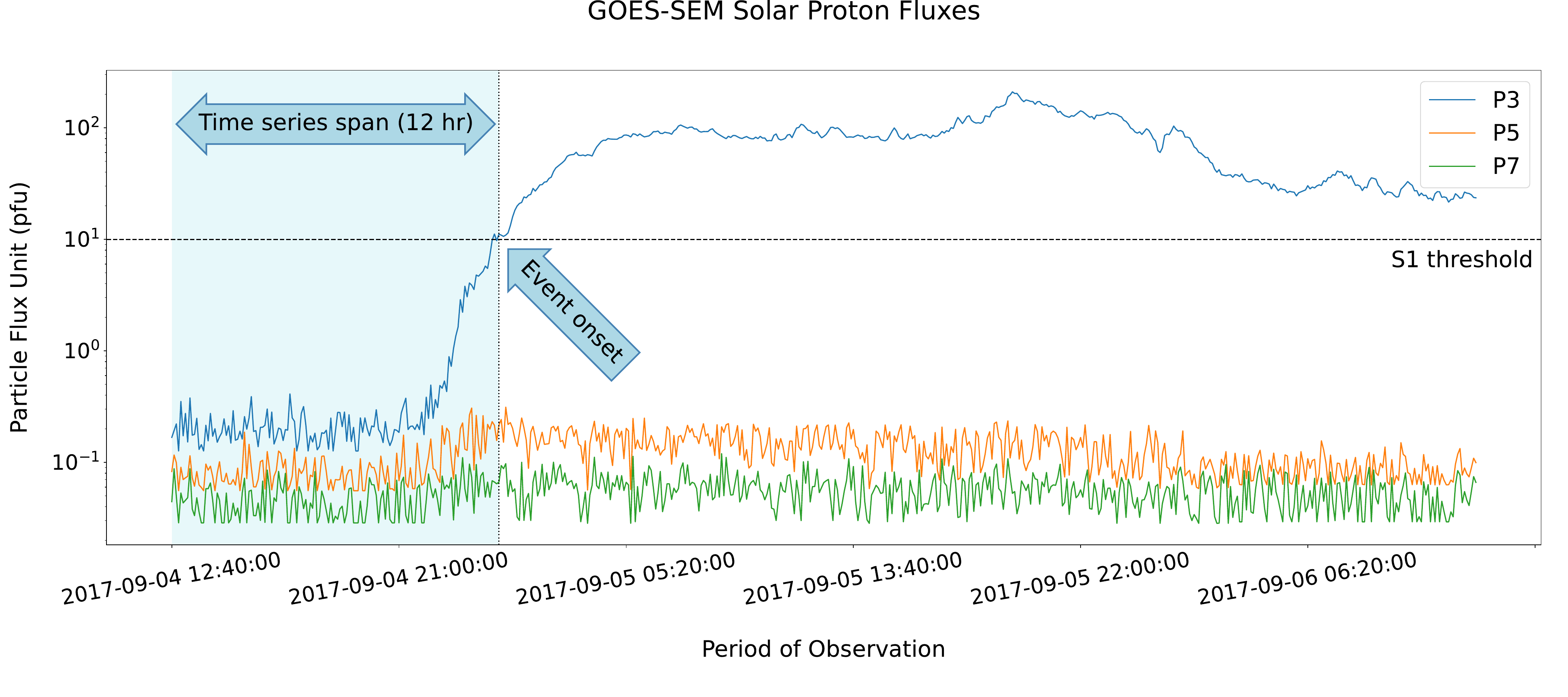}
\caption{Time series plot of a large SEP event that occurred on 2017-09-05T00:40 (UT) shown on a log scale that reached a peak proton flux of $\approx$210 pfu on 2017-09-05T19:30 (UT). The three fluxes in the legend correspond to GOES P3 (E$\geq$10 MeV), P5 (E$\geq$50 MeV) and P7 (E$\geq$100 MeV) integral proton channels. The horizontal black dashed line indicates the SWPC threshold for a large SEP event while the vertical black dotted line indicates the SEP event onset time. The shaded region shows the typical span of time series considered in our work. It corresponds to 12 hours of proton fluxes prior to the SEP event onset.
\label{fig:ts}}
\end{figure}

As reported by Rotti \& Martens (\citeyear{rotti2023analysis}), $\approx$ 79\% of SEP events have a precursor eruption within 12 hr prior. In other words, most SEP events' onset times are within 12 hr after the initiation of the parent flare eruption. Interestingly, most (53) events with a parent eruption more than 12 hr prior to SEP onset occur during solar maximum ($\pm$ one year). Many of these precursor eruptions occur more than a day before the onset of an SEP event. We consider 12 hr as an optimal span or observation window in the present work. However, limiting the observation window to 12 hr does not cause a huge limitation on our models. That is, the inclusion of X-ray fluxes is valuable but not trivial to short-term predictions of SEP events. Hence, we have considered 12-hr as an optimal window by including as much precursor (X-ray) data as possible. Increasing the window length to greater than 12-hr has the potential to induce noise such as additional and unrelated X-ray flux peaks in data. In addition, we omit five minutes of input data just before the SEP event onset. As we consider fluxes with \edit1{one}-minute cadences, our data set represents a 715-length soft X-ray and integral proton time series. A sample time profile for a large SEP event in the GSEP data set is shown in Figure \ref{fig:ts} that occurred on 2017-09-05T00:40 (UT) with a rise time of $\approx$ nineteen hours. The parent flare erupted about four hours before the SEP event onset from active region 12673 (solar lon = 12\degree, solar lat = -10\degree) and had a magnitude of M5.5 as measured by the GOES/XRS instrument. Following the flare there was a halo fast-CME propagating with a velocity of $\approx$1400 km.s$^{-1}$. The SEP event reached a peak flux of $\approx$210 pfu on 2017-09-05T19:30 (UT) in the E$\geq$10 MeV channel measured by the GOES-SEM instrument. The vertical dotted line overlayed in the plot indicates the event's start time while the horizontal dashed line indicates the SWPC S1 threshold. The shaded region shows the typical length of the time profile we utilize in our work.

\subsection{Data Labels}
The work discussed here considers the term `SEP events' analogous to solar protons events (SPEs). While variations exist, event labels are usually associated with the occurrence of strong/large SEPs based on the integral proton fluxes ($I_{P}$) recorded by P3 crossing the 10 pfu threshold. As mentioned earlier, the small or sub-events are defined based on a threshold of 0.5$>$$I_{P}$$<$10 pfu in the 10 MeV channel. If there are successive SEP events within 12 hours, then the observation window shall constitute fluxes prior to the former event onset. There are several events reported in the GSEP data set that have overlapping proton fluxes from the previous event. Due to the nature and characteristics of the SEP event, such overlapping cannot be excluded. In these scenarios, when the proton fluxes in the 10 MeV channel are already above 10 pfu, the model outputs a “yes” label indicating a large event. This ``back-to-back events'' situation is evident during solar maximum. In the GSEP data set, 23 (4) large (small) SEP events occur within the next 24 hours following the first event. There are only six successive events occurring within 12 hours, all of which are large in nature, with a median rise time of $\approx$14 hours and a median event length of $>$48 hours.

Another critical threshold in terms of operational requirements concerning astronauts during extra-vehicular activities is one pfu in the E$\ge$100 MeV (P7) channel. Nonetheless, in the present work, we focus only on the SWPC `S1' threshold and defer the former scenario to future work.

In the context of solar particle radiation a passing interplanetary shock causes energetic storm particle (ESP) acceleration (Cane \citeyear{CANE199535}). Although ESPs are different kinds of particle events, they can still be brought under the ``umbrella" of SEPs since the energetic particle fluences still determine the radiation exposure and dosage rate. Furthermore, it is relevant to minimize the total dosage rate of an astronaut during a space mission for their health and safety. Therefore, our focus has been a cumulative ``solar particle event" prediction wherein we also include the nine ESPs reported in the GSEP catalog in our analysis.

\section{Methodology} \label{sec:methods}
In this work, we attempt to address the grand problem of SEP event predictions from a time series classification perspective. This problem is constructed here in the framework of a binary classification task. Here, the target labels are based on surpassing the proton flux threshold defined by NOAA-SWPC. Accordingly, the SEP event class labels that have proton enhancements above the threshold ($I_{P}$$\geq$10 pfu) are ``positive", else ``negative". In this section, we describe a novel framework for classifying E$\geq$10 MeV SEP events using time series-based ML models.

\begin{figure}[ht!]
\plotone{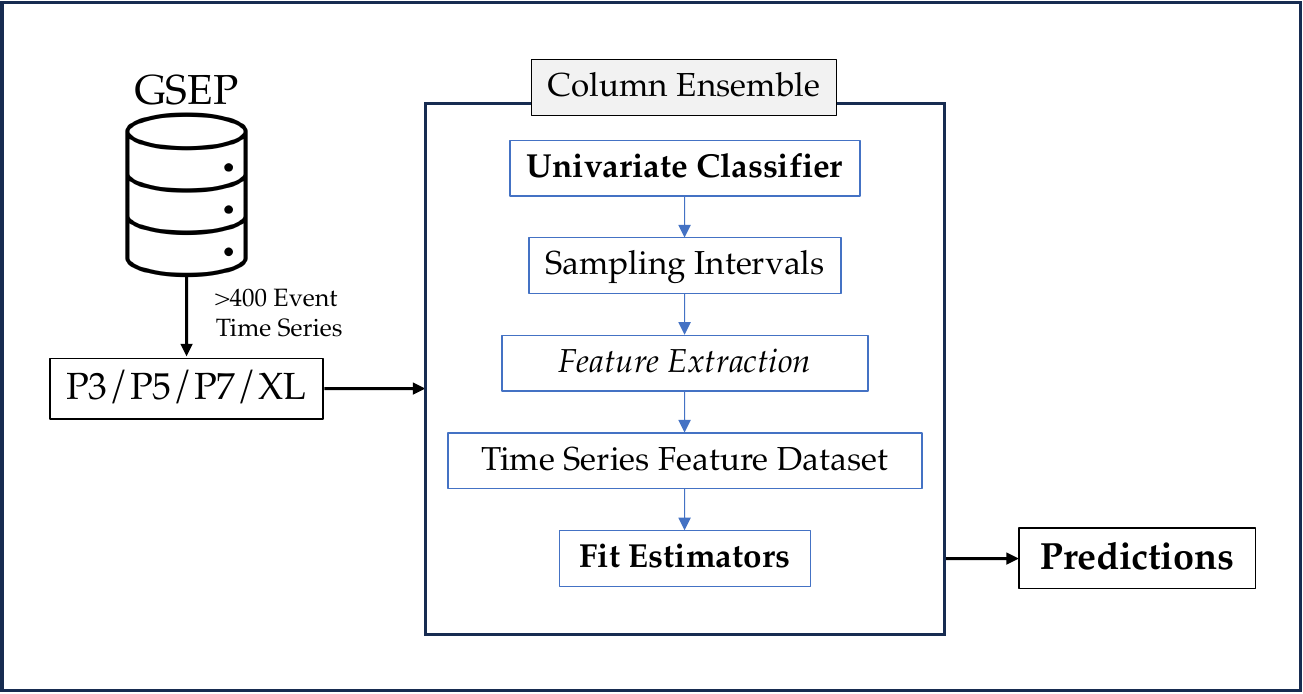}
\caption{Schematic overview of the workflow. We consider three proton and the long band X-ray channels from the GSEP time series data set. In a column ensemble, we input the fluxes to our classifiers. Each univariate classifier subsamples the time series and extracts features from each interval to generate a feature data set. The classifier is trained to fit the input data and further tested on unseen data.
\label{fig:schema}}
\end{figure}

We use a column ensemble of univariate classifiers, a parameter-wise ensemble of columns in which
individual classifiers are applied to every parameter (column). This is a homogeneous ensemble schema; an overview of it is shown in Figure \ref{fig:schema}. The ensemble estimator allows multiple feature columns of the input to be transformed separately. The statistical features generated by each classifier on samples of the original time series are ensembled to create a single output. Each feature is assigned a score that indicates how informative it is towards predicting the target variable (Hansen \& Salamon \citeyear{hansen1990neural}; Schapire \citeyear{schapire1990strength}; Arbib \citeyear{arbib2003handbook}). 

In our case of the GSEP data, we create a multivariate variant of univariate algorithms using the column ensemble method described above. We consider the long band X-ray (xl) and three proton channels (P3, P5, P7) as our input time series. We implement and compare the performances of three classifiers for large/small SEP event classifications. The prediction results from these individual column classifiers are then aggregated as a whole (with equal votes using prediction probabilities). The idea is to see if the observed time series span leads to a large SEP event (positive class) or not (negative class). The negative classes here do not constitute SEP-quiet periods but are entirely small SEP events. These sometimes behave almost as large events but fall below the critical threshold. Identifying such patterns is relevant to reducing false alarms. In other words, the reason for choosing these two classes is that the models must pick up the incoming flux behavior of X-rays and earthward accelerating protons that may cross the SWPC event threshold, which requires mitigation measures in an operational context. The rationale is to explore the operationally relevant proton channels including those with the xl channel.

Regarding existing SWx forecasting methods, flare forecasters build models distinguishing between $\geq$M1.0 and $\leq$C9.9 classes (Ji et al. \citeyear{ji2020all}). Similarly, we aim to provide an interpretable state-of-the-art time series ML model to classify large and small SEP events. Therefore, this method will provide a perspective to extend the univariate time series classifiers in an ensemble and build a prototype short-term SEP event prediction system that optimizes the model based on forecast skill scores. Section \ref{sec:tsc} provides more details about these classifiers and their feature sets.

\subsection{Time Series Classification}\label{sec:tsc}
SWx practitioners and forecasters highly recommend using temporal features and time series analysis for better forecasting (Singer et al. \citeyear{singer2001space}). In time series data, every timestamp is typically a vector or array of real values observed over time. It can be divided into univariate or multivariate such that an array of only one parameter is a univariate series and a set of univariate series forms a multivariate series (Ruiz et al. \citeyear{ruiz2021great}). In time series-based ML, one of the techniques to improve model performance is the reduction of the dimensionality of the data set by identifying and choosing the most relevant features (Koegh et al. \citeyear{keogh2001dimensionality}; Cassisi et al. \citeyear{cassisi2012similarity}).

Feature-based models extract highly relevant statistical features from the time series that are later used as a core subset in training  models (Fulcher \& Jones \citeyear{fulcher2014highly}). This step has multiple purposes, such as (1) optimizing the performance of the models by choosing relevant features, (2) providing robust predictors thereby reducing computational costs, and (3) offering better interpretability to the underlying physical processes that generated the data model. Time series classification uses supervised ML to analyze labeled classes of time series data and then to predict the class to which a new data set belongs. This is important in SWx predictions, where particle sensor data is analyzed to support operational decisions in near-real-time (NRT). The accuracy of classification is critical in these situations, and hence we must ensure that the classifiers are as accurate and robust as possible.

There are many algorithms that are designed to perform time series classification. Depending on the data, one type might produce higher classification accuracies than other types. This is why it is important to consider a range of algorithms when considering time series classification problems. In this work we experiment with interval-based and dictionary-based models on our data set.

Interval-based algorithms typically split the time series into multiple random intervals. Each temporal feature calculated over a specific time series interval can capture some essential characteristics. Therefore, the algorithm gathers summary statistics from each sub-series to train individual classifiers on their interval. Next, the most common classes are evaluated among the intervals and return the final class label based on equal voting for the entire time series (Bagnall et al. \citeyear{bagnall2017great}).

On the other hand, dictionary-based models implement the bag of words (Zhang et al. \citeyear{zhang2010understanding}) algorithm. In a broad structure a sliding window of length `l' runs across a series of length `n'. Then, all real-valued window lengths are converted into a symbolic string called a ``word'' through approximation and discretization processes. During this process, the possible representations are stored in a dictionary. At the end of the series length, the occurrence of each ``word'' from the dictionary in a series is counted and transformed into a histogram. Finally, histograms of the extracted words are used for the classification task of new input data (Faouzi \citeyear{faouzi2022time}).

Amongst the univariate interval-based approaches, we consider Time Series Forest (TSF; Deng et al. \citeyear{deng2013time}) and Supervised Time Series Forest (STSF; Cabello et al. \citeyear{cabello2020fast}). From dictionary-based classifiers, we use the Bag of SFA Symbols (BOSS; Sch{\"a}fer \citeyear{schafer2015boss}) that uses the Symbolic Fourier Approximation (SFA; Sch{\"a}fer \& H{\"o}gqvist \citeyear{schafer2012sfa}) to transform and discretize subseries into words. We explain the model structure below. A brief summary of the model functions and parameters is presented in Table \ref{tab:props}. All our computational experiments are performed using the Python programming language (Sanner et al. \citeyear{sanner1999python}). All the classifiers used in this study are from the sktime library (L{\"o}ning et al. \citeyear{markussktime}).

\begin{deluxetable*}{lcc}[ht!]
\tablenum{1}
\tablecaption{Summary properties of the models. \label{tab:props}}
\tablewidth{0pt}
\tablehead{
\textbf{Model} & \textbf{Sampling schema} & \textbf{Features}}
\startdata
TSF  & Random intervals & $\mu$, $\sigma$, \textit{m} \\
STSF & Supervised intervals & $\mu$, $\sigma$, \textit{m}, median,\\
  &  & IQR, min, max  \\
BOSS & Sliding window   & Word representations \\
\enddata
\tablecomments{
Model names: TSF - Time Series Forest; STSF - Supervised Time Series Forest; BOSS - Bag of Symbolic Fourier approximation Symbols. --- Feature names: $\mu$ - Mean; $\sigma$ - Standard deviation; \textit{m} - Slope; IQR - Interquartile range; min - Minimum value; max - Maximum value.
}
\end{deluxetable*}

\begin{figure*}[hb!]
\gridline{\fig{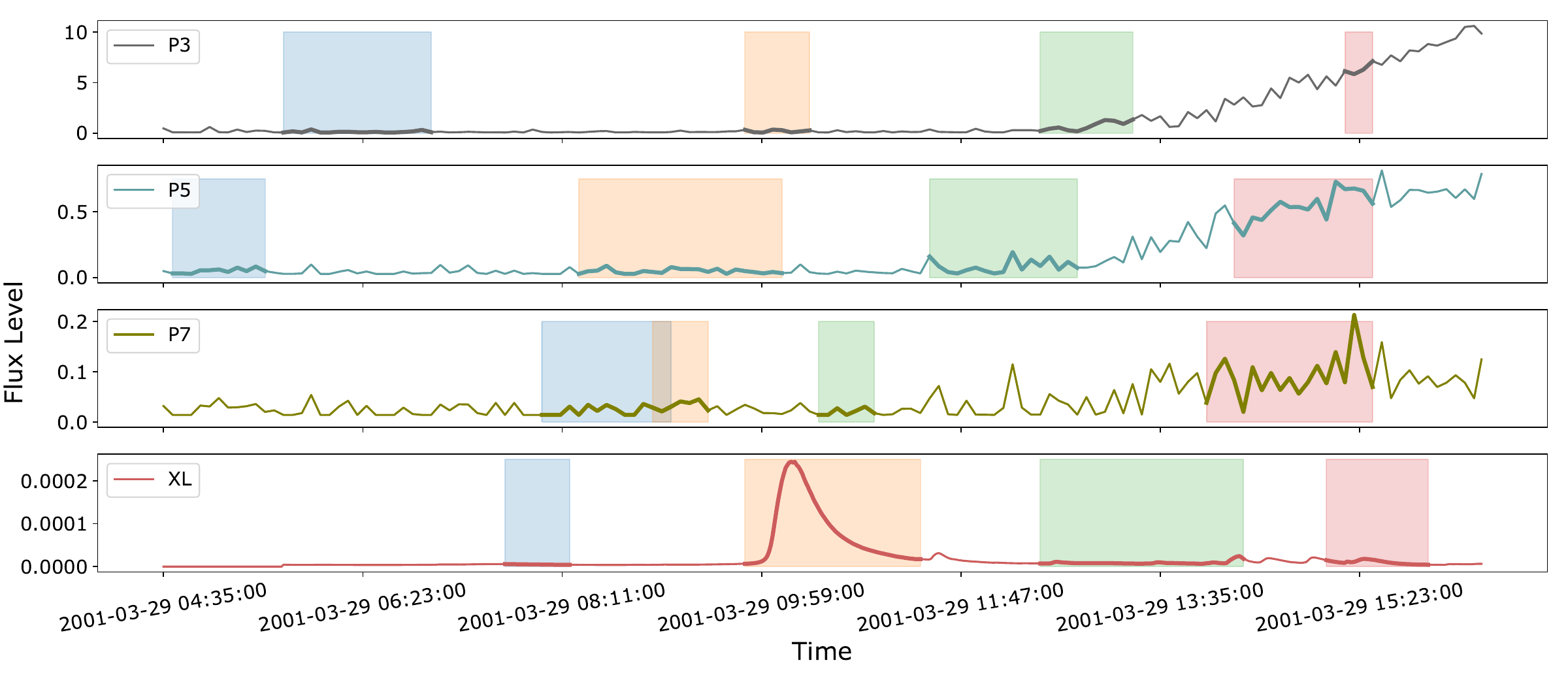}{0.7\textwidth}{(a)}
          }
\gridline{\fig{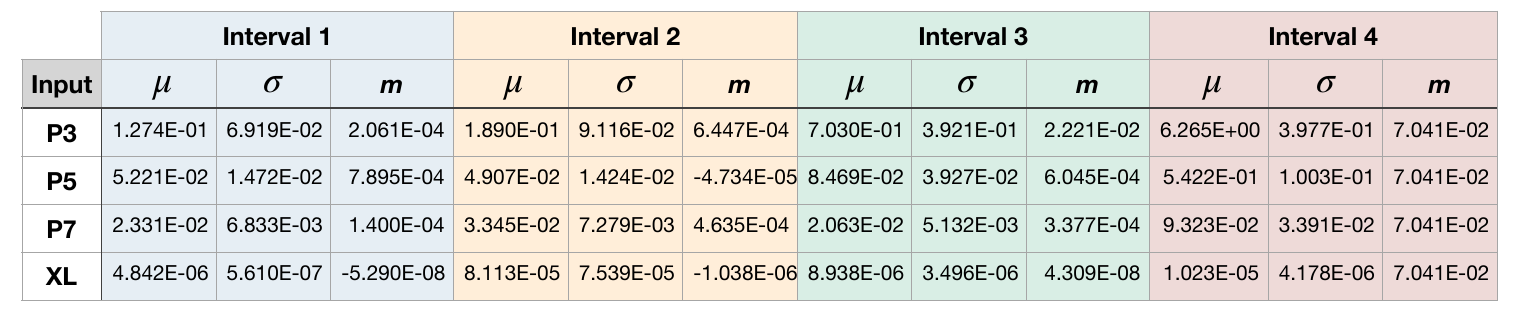}{0.7\textwidth}{(b)}
          }
\caption{Schematic overview of time series forest (TSF) model. (a) Random intervals are generated and the corresponding subsets from each time series are extracted. (b) Three statistical features are derived from each subintervals: mean ($\mu$), standard deviation ($\sigma$) and slope (m).}
\label{fig:tsf}
\end{figure*}

\subsubsection{Time Series Forest}
One of the most commonly used and popular interval-based algorithms is the time series forest (TSF; Deng et al. \citeyear{deng2013time}). This model implements a random forest approach where multiple decision trees are grouped. Each tree in this ensemble is trained using a subset of statistical features derived from randomly selected intervals, essential in reducing the dimensionality of high-dimensional feature spaces. The statistical features derived from random intervals are mean ($\mu$), standard deviation ($\sigma$) and slope of the regression line (m). Figure \ref{fig:tsf} illustrates the feature extraction process from random intervals in the TSF algorithm. The process of obtaining statistical summaries of intervals is called flattening the vectors. Each decision tree classifier then assigns a target label to its interval of the data based on a majority vote of all trees. The voting process is needed since every single tree only evaluates a certain subseries of the time series.

\subsubsection{Supervised Time Series Forest}
\begin{figure}[hb!]
\plotone{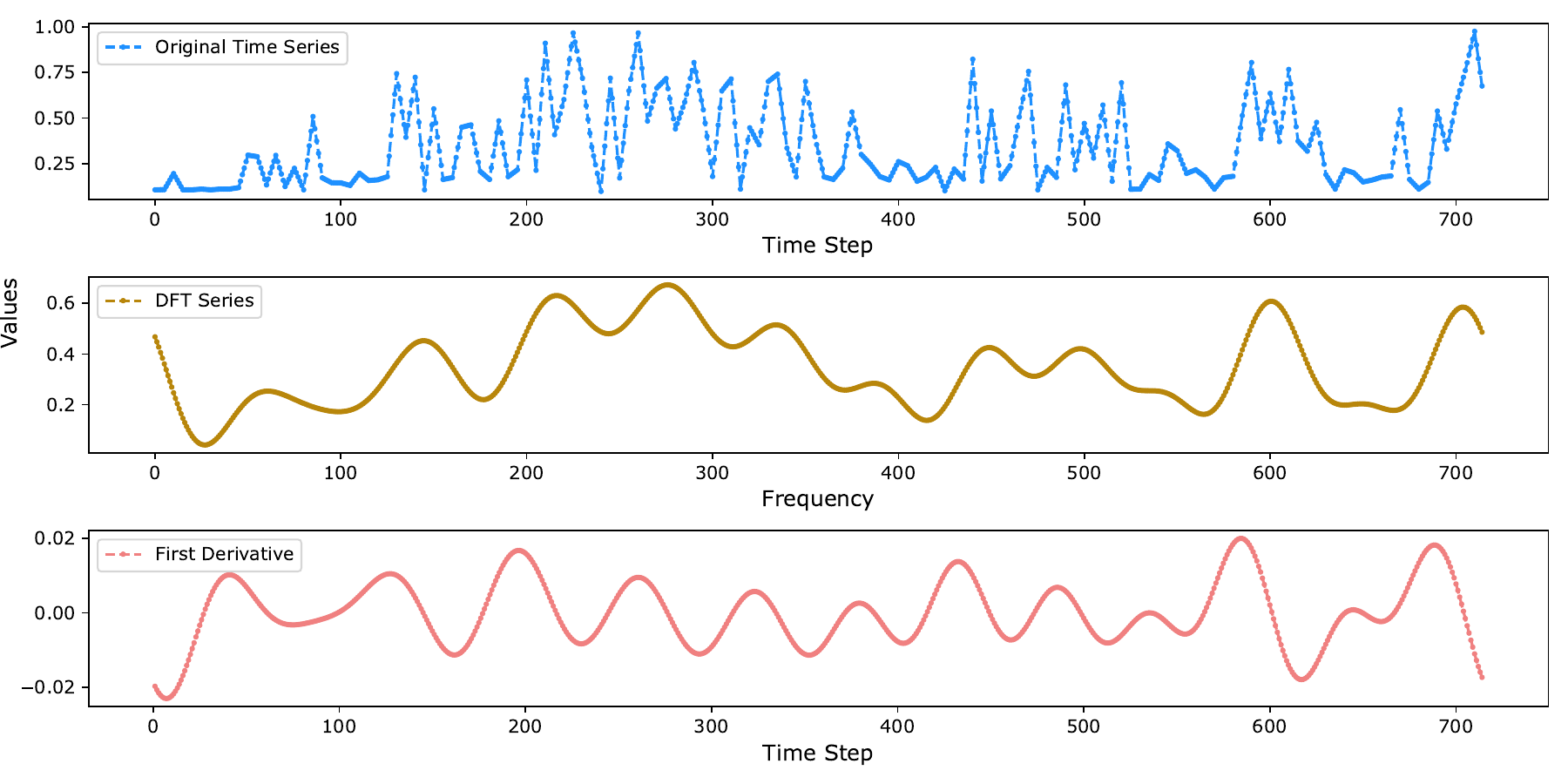}
\caption{Schematic overview of time series representations of STSF. For a given original series, a periodogram representation derived from the discrete Fourier transom (DFT) and a first-order difference representation are generated to find candidate discriminatory intervals as a subset of the original time series. The discriminatory interval features constituting seven statistical parameters are obtained from all three (time, frequency, and derivative) domains prior to training the classifier.}
\label{fig:stsf}
\end{figure}

Another interval-based model is the supervised time series forest (STSF; Cabello et al. \citeyear{cabello2020fast}). Here, an ensemble of decision trees is built on intervals selected through a supervised process wherein the algorithm finds the discriminatory intervals. The ranking of the interval feature is obtained by a scoring function that indicates how well the feature separates a class of time series from the other classes. The final set of intervals is obtained in a top-down approach to represent the entire series. STSF aims to improve the classification efficiency by selecting in a supervised fashion (based on their class-discriminatory capabilities) only a subset of the original time series. The algorithm uses three (time, frequency and derivative) representations of the time series as shown in Figure \ref{fig:stsf} and extracts seven features ($\mu$, $\sigma$, m, median, interquartile range (IQR), minimum value and maximum value) from each interval. Finally, the feature set is concatenated to form a new data set upon which decision trees are built. The final output is based on majority voting of averaged probability estimates of the ensemble.

\subsubsection{Bag-of-SFA-Symbols }
\begin{figure*}[ht!]
\gridline{\fig{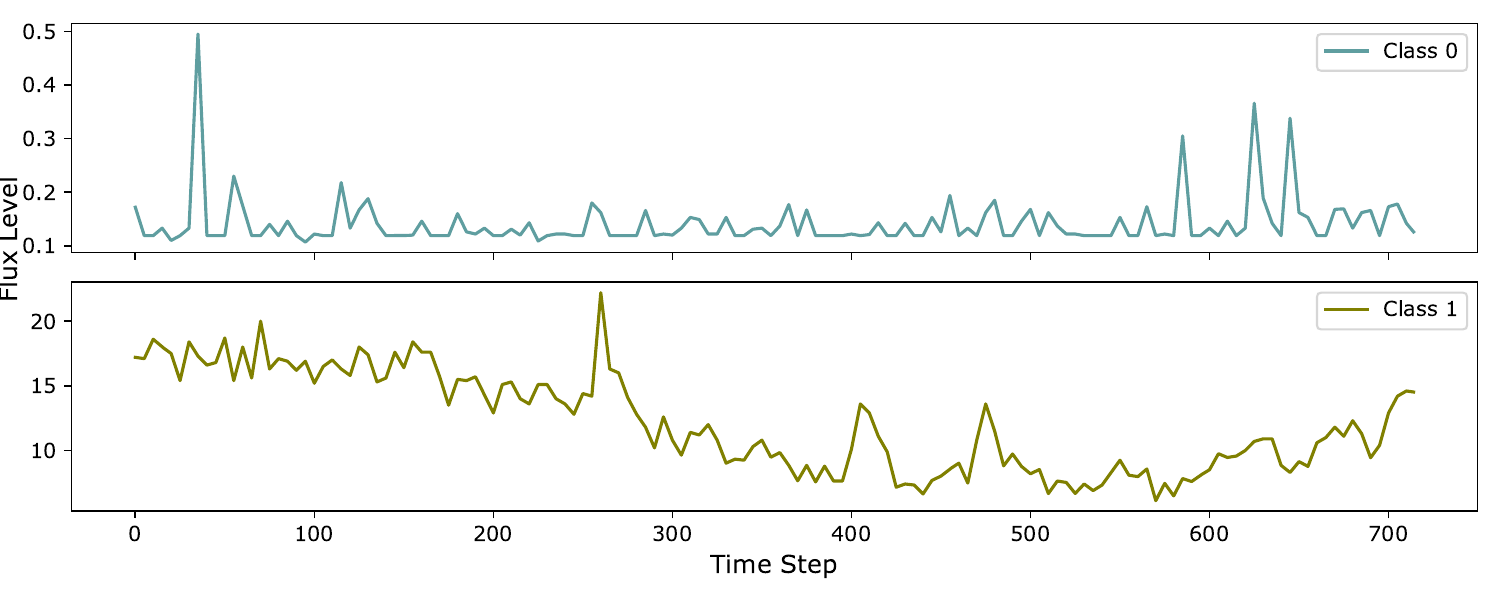}{0.8\textwidth}{(a)}
          }
\gridline{\fig{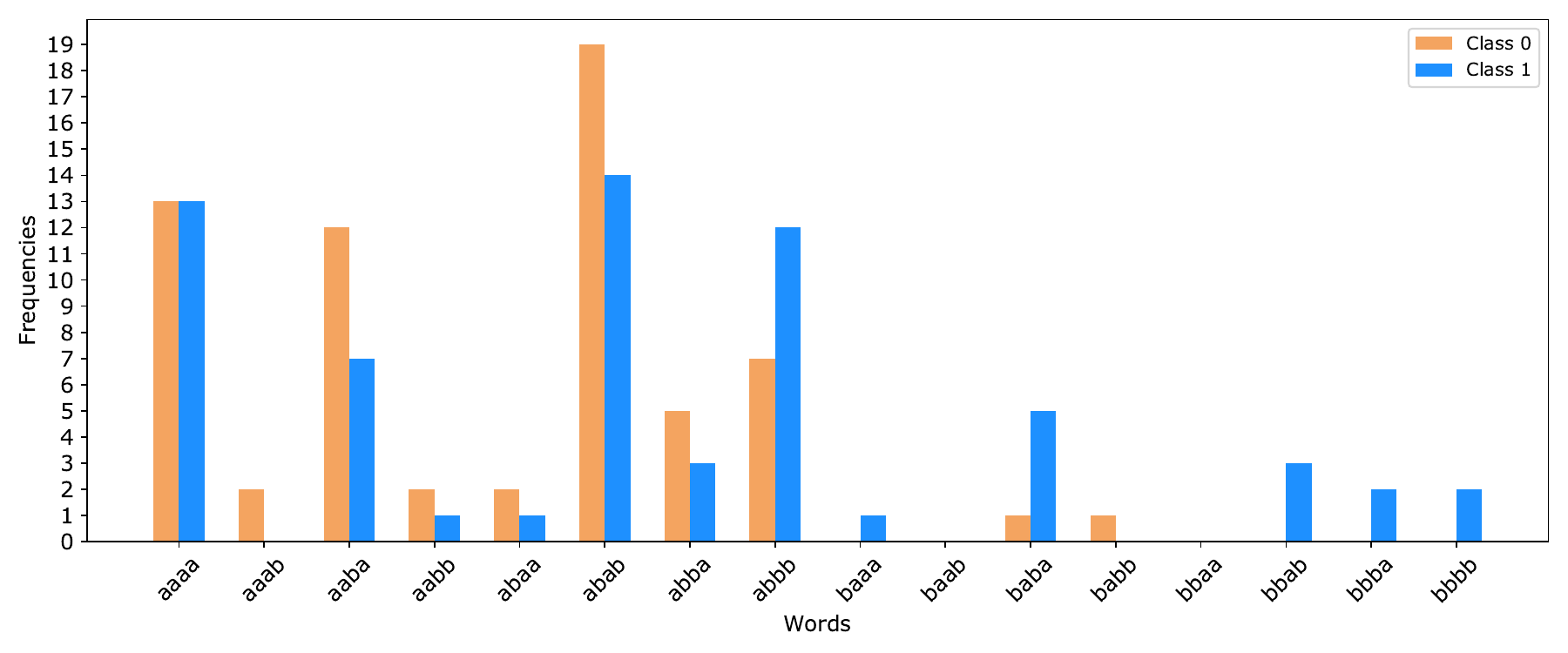}{0.8\textwidth}{(b)}
          }
\caption{Schematic overview of bag-of-SFA symbols (BOSS) model. (a) Given a raw time series, a sliding window is applied to extract subsequences. Each subsequence is transformed into a word using the symbolic Fourier approximation (SFA) algorithm, and only the first occurrence of identical back-to-back words is retained. (b) Lastly, a histogram of the words is computed.}
\label{fig:boss}
\end{figure*}

The bag of symbolic Fourier approximation symbols or BOSS algorithm (Sch{\"a}fer \citeyear{schafer2015boss}) typically uses a sliding window to transform the time series into sequences of symbols to extract ``words" and form a histogram. The final classification is made by determining the distribution of these ``words'' in the histogram. The intuition behind this method is that times series are similar, which means they are of the same class if they contain similar ``words''. Firstly, BOSS finds symbolic approximations using discrete Fourier transform (DFT). Then, it creates words and discretizes/vectorizes the input using words with multiple coefficient binning (MCB). This has the effect of reducing noise (Sch{\"a}fer \citeyear{schafer2015boss}). Finally, the algorithm uses a one-nearest neighbor over word frequency vectors and retains the estimators using the BOSS metric for best parameter training (Bagnall et al. \citeyear{bagnall2017great}). Figure \ref{fig:boss} illustrates these stages of the BOSS algorithm.

\subsection{Data Partitions}
For classification in a supervised setting where all the data has class labels, the data set is typically split into the training set and the test set (Hastie et al. \citeyear{hastie2009elements}). The training set is used to fit the data features on the parameters of the algorithms chosen to address the problem. The chosen algorithm is used to score the test set and determine the quality of the classifier. We partition our data into training-test sets with splitting criteria of 65-35 percent leading to 283 training samples and 150 test samples. A summary of the number of samples in each partition with respect to the target labels is presented in Table \ref{tab:dp}.

\begin{deluxetable*}{r|c|c|}[ht!]
\tablenum{2}
\tablecaption{Data partitioning. \label{tab:dp}}
\tablewidth{0pt}
\tablehead{
\textbf{} & {\textbf{Training}} & {\textbf{Test}}}
\startdata
Positive & 167 & 77 \\
|||| &|||&|||\\
Negative & 116 & 73 \\
\enddata
\tablecomments{Number of instances in each partition corresponding to the binary target labels. Here, binary corresponds to a positive (strong SEP event) or negative (weak SEP event).}
\end{deluxetable*}

\section{Results} \label{sec:results}
In this work, we consider large ($\geq$ S1) SEP events as a `positive' class and small events as a `negative', thereby designing the problem as a binary classification task. The experiments are designed to fit a univariate model to a multivariate time series architecture for a short-term SEP event prediction system. We aim to demonstrate the robustness and compare the efficiency of time series classifiers towards generating short-term predictions during NRT operations. As explained in the previous section, the classifiers extract the features and data attributes from the input series. Because we want to aim at short-term predictions via SEP event classification, we consider 12 hours of observations minus five minutes before the SEP event onset. Here, the onsets are defined as follows: large events crossing 10 pfu and small events surpassing 0.5 pfu in the P3 channel. We interpolate the five-minute proton time series to one minute to utilize the X-ray flux characteristics during flaring periods. The model hyperparameters considered are as follows: (i) Minimum interval length/window size is fifteen for TSF and BOSS, and (ii) Number of estimators is 200 for TSF and STSF. 

\subsection{Learning Curves}\label{subsec:lc}
One of the essential tools in ML to trace the model performance is using learning curves. These curves visually indicate the sanity of a model for overfitting or underfitting during the training phase. They also help us to understand how the model performance changes as we input more training examples. In addition, these curves are useful to compare the performance of different algorithms (Perlich et al. \citeyear{perlich2003tree}).

\begin{figure*}[b!]
\gridline{\fig{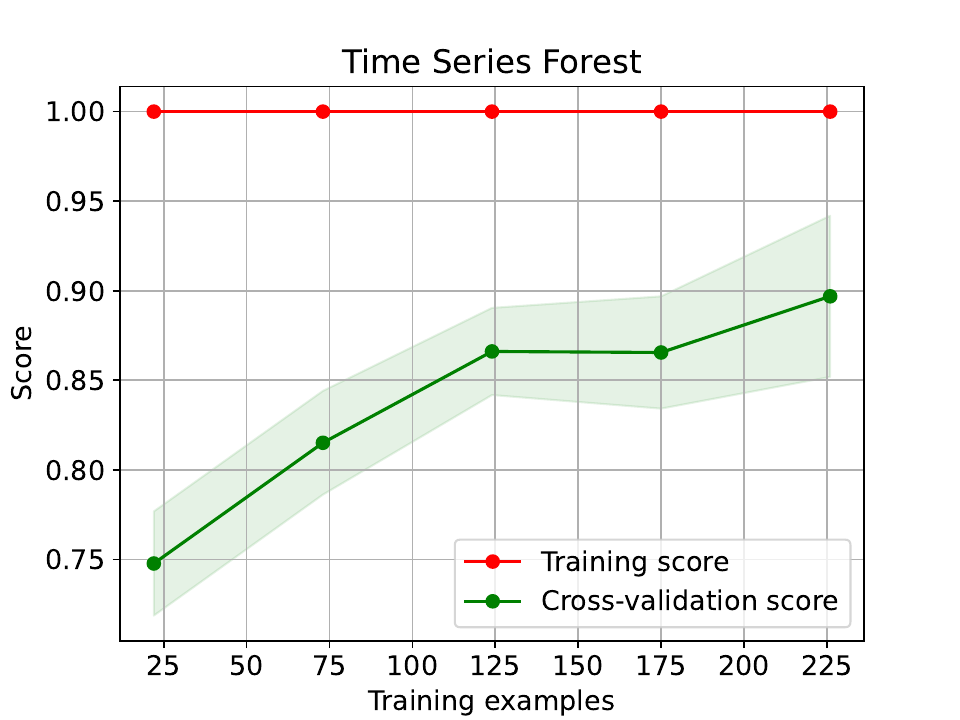}{0.49\textwidth}{(a)}
          \fig{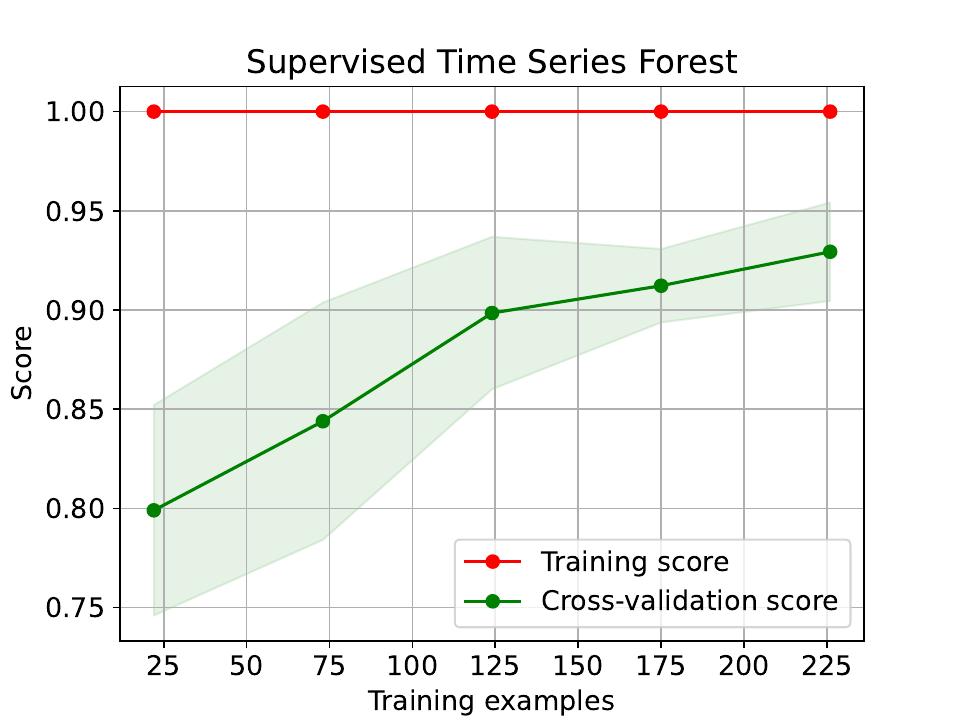}{0.49\textwidth}{(b)}
          }
\gridline{\fig{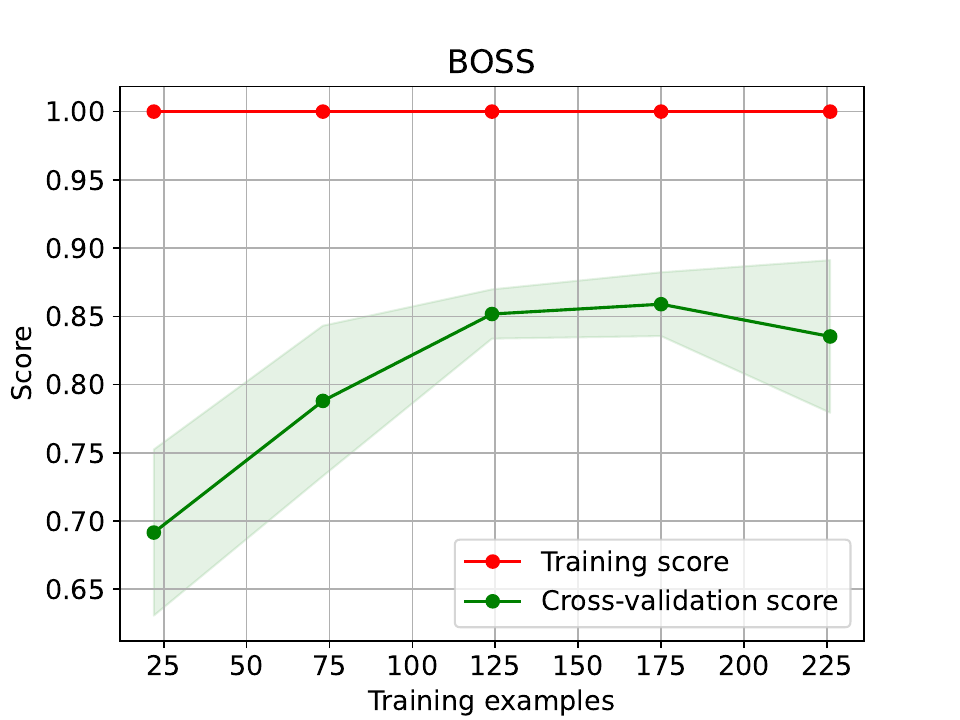}{0.49\textwidth}{(c)}
          }
\caption{Learning curves for (a) time series forest (TSF); (b) supervised TSF (STSF); and (c) BOSS ensemble models. The weighted F$_{1}$-score has been used here as the scoring function. The red line represents the training score, while the green line shows the model estimations on validation. Here, the shaded region indicates the standard deviation of the validation score. The STSF model produces the best score ($\approx$0.925) at the end of cross-validation.
\label{fig:lc}}
\end{figure*}

Figure \ref{fig:lc} shows the learning curves of the models in our consideration. Here, for providing a better performance estimate given the imbalanced nature of our data set, we use a `weighted' average of F$_{1}$-scores (Manning et al. \citeyear{manning_raghavan_schütze_2008}) per class as defined in Equation \ref{sc:f1_wt}.

\begin{equation} \label{sc:f1_wt}
F_{1_{weighted}} = \sum_{i=1}^{N} w{_i} \times F_{1_i}
\end{equation}

\begin{equation} \label{sc:f1}
F{_1} = 2 \times \frac{(Precision \times Recall)}{(Precision + Recall)}
\end{equation}

As shown in Equation \ref{sc:f1}, F$_{1}$-score can be estimated as the harmonic mean of precision (Eq.\ref{sc:pre}) and recall (Eq.\ref{sc:rec}). Precision is used to evaluate the model's correct prediction with respect to the false alarms. Recall characterizes the ability of the classifier to find all of the positive cases.

\begin{equation} \label{sc:pre}
Precision = \frac{(TP)}{(TP + FP)}
\end{equation}

\begin{equation} \label{sc:rec}
Recall = \frac{(TP)}{(TP + FN)}
\end{equation}

As we consider a `weighted' average for the F$_{1}$-score, it computes the score for each target class and uses sample weights that depend on the number of instances in that class while averaging. The weight in the F$_{1}$-score is presented in Equation \ref{sc:wt}. Here, i is the number of target classes in the data set, which is two in the present work.

\begin{equation} \label{sc:wt}
w{_i} = \frac{\text{Number of samples in class \textit{i}}}{\text{Total number of samples}}
\end{equation}

In our learning curves, the red line represents the training score, which evaluates the model on the newly trained data. The green line shows the estimations of the model on the samples used for validation. The shaded area represents the standard deviation of the scores after running the model multiple times with the same number of training data. It can be seen that the training score remains high for all models regardless of the size of the training set.

In Figure \ref{fig:lc}(a), the steepness of the green line reaches a plateau between $\approx$ 125 to 175 samples but shows a small increment after 175 for TSF. On the other hand, the cross-validation score in Figure \ref{fig:lc}(b) for STSF greatly reduces after 125 samples. In Figure \ref{fig:lc}(c), the curve for BOSS model initially increases with the training size up to $\approx$125 but the slope reduces later, indicating that more training data is not helpful in the generalization process. The STSF model achieves a high F$_{1}$-score ($\approx$ 0.925) followed by TSF and then the BOSS model. Overall, the learning curves represent a satisfactory use of sample sizes to train the model efficiently. For TSF and STSF, we note that with more samples, this can be improved. In the remainder of this section, we present and discuss the implementation of several evaluation techniques to analyze the performances of the models on the test set.

\subsection{Reliability Curves}
In ML, reliability curves/calibration plots are used to better understand a model's confidence intervals in its prediction probabilities. Models such as decision trees give the label of the event but do not support native confidence intervals. A simple decision tree is a hierarchical tree structure used to determine classes based on a set of rules (questions) about the attributes of the data points (Safavian \& Landgrebe \citeyear{safavian1991survey}). Here, every non-leaf node represents an attribute split (question), while all the leaf nodes represent the classification result. In short, if the decision tree model is input with a set of features and corresponding classes, it generates a sequence of criteria to identify a data sample's target class. 

\begin{figure}[ht!]
\plotone{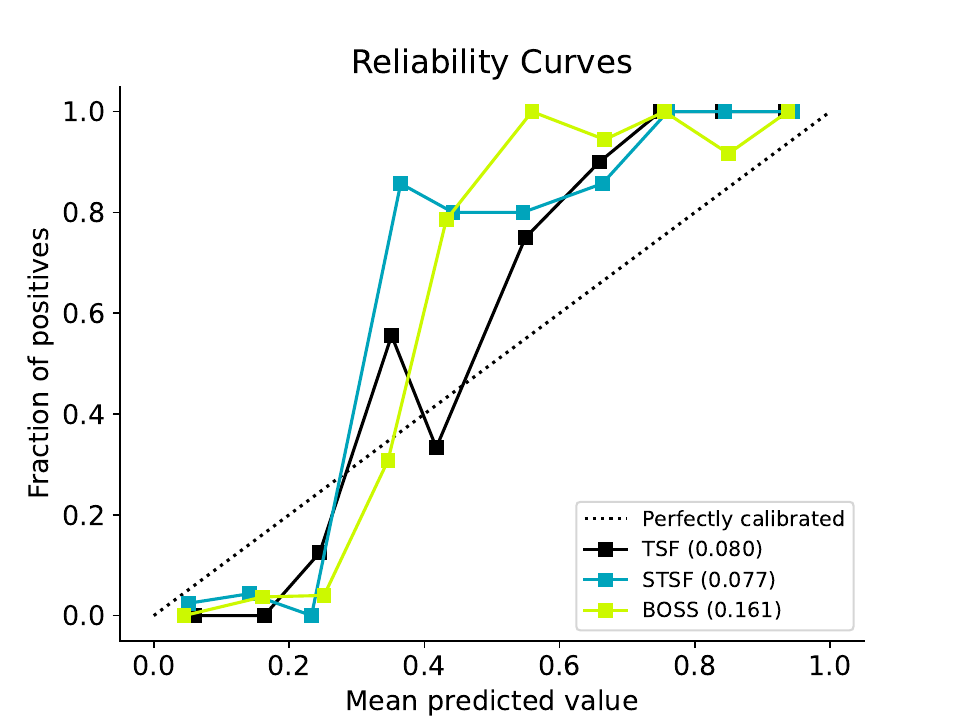}
\caption{Reliability diagram or calibration plots of our models on the test set. The diagonal black dotted line shows the best fit. Data points above this line are underestimates, while those below it are overestimates. Shown in the legend are model names; TSF, STSF and BOSS with Brier score loss, respectively.
\label{fig:calib}}
\end{figure}

We can evaluate the models based on multiple tools to be confident in our predictions. One method is calibration plots that check whether the predicted class distributions are similar to the true ones. Calibration curves (Wilks \citeyear{wilks1990combination}) visually aid us in comparing how well the probabilistic predictions of a binary classifier are calibrated. Figure \ref{fig:calib} shows the predicted probability of a model in each bin on the x-axis and the fraction of the positive label in that bin on the y-axis. The calibration intercept seen in a black-dotted line is a best-fit assessment. Values under the curve suggest overestimation, whereas values above the curve suggest underestimation.

TSF and STSF show close behavior in their average predictions over true values compared to the BOSS model. Nonetheless, all the models show underestimates of their predictive probability against the observed probability. In other words, this represents relatively lower confidence intervals in the model predictions. Hence, we use the Brier score (BS) loss (Murphy \citeyear{murphy1973new}) as defined in Equation \ref{sc:bs} to evaluate the performance of the model.

\begin{equation} \label{sc:bs}
    BS = \frac{1}{N} \sum_{i=1}^{N} (y_{i} - \hat{y}_{i})^{2}
\end{equation}

Here, N is the number of data samples in the test set; $y_{i}$ is the observed probability and $\hat{y}_{i}$ denotes the prediction score (used as the estimated probability) of the $i^{th}$ test sample. Brier score loss is strictly used to assess the calibration and discriminative power of a model, as well as the randomness of the data at the same time. The loss values range from 0 to 1, with 0 being a perfect score. In our case, TSF has \textbf{0.080}, STSF has \textbf{0.077} and BOSS has \textbf{0.161} as Brier score losses. Because of the low losses, our models indicate they are excellent predictors with more discriminatory power. Therefore, we further evaluate the model on the test set using popular metrics and compare their performances.

\subsection{Evaluation}
In Section \ref{subsec:lc}, we have defined statistical metrics, such as precision and recall, that have been traditionally used to assess classifier performances. On a simple scale, accuracy (Eq.\ref{sc:acc}) is another standard evaluation metric used to evaluate the quality of a classifier by counting the ratio of correct classification over total classifications.

\begin{equation} \label{sc:acc}
Accuracy = \frac{(TP + TN)}{(TP + FP + TN + FN)}
\end{equation}

Furthermore, we can focus on false negatives and measure the model performance using a receiver operating characteristic (ROC) curve. The ROC curve for the classifier is generated by plotting the true positive rate (TPR) against the false positive rate (FPR). The classifier predicts mean probabilities for each input instance belonging to the positive class, where the prediction score from the classifier is greater than a parametrized threshold. Then, a classification threshold (in the range 0 to 1) is used to assign a binary label to the predicted probabilities. To find the optimal threshold that minimizes the difference between TPR and FPR of the classifier, we use the Youden Index (J; Youden \citeyear{youden1950index}) defined in Equation \ref{sc:J}. Here, sensitivity is the recall for the positive class and specificity is the recall for the negative class. We further explain our analysis on finding the optimal threshold in Appendix \ref{sec:TA}.

\begin{equation} \label{sc:J}
J = Sensitivity + Specificity - 1
\end{equation}

The quality of the model is then assessed on the area under the ROC curve (AUC) for the positive class. The intuition behind this measure is that AUC equals the probability that a random positive sample ranks above a random negative sample. Ahmadzadeh et al. (\citeyear{ahmadzadeh2019challenges}) point out that the AUC is statistically consistent and more discriminating than accuracy. A measure of 1.0 for AUC signifies perfect classification, while a value of 0.5 means that the classifier cannot differentiate at all. 

\begin{figure*}[ht!]
\gridline{\fig{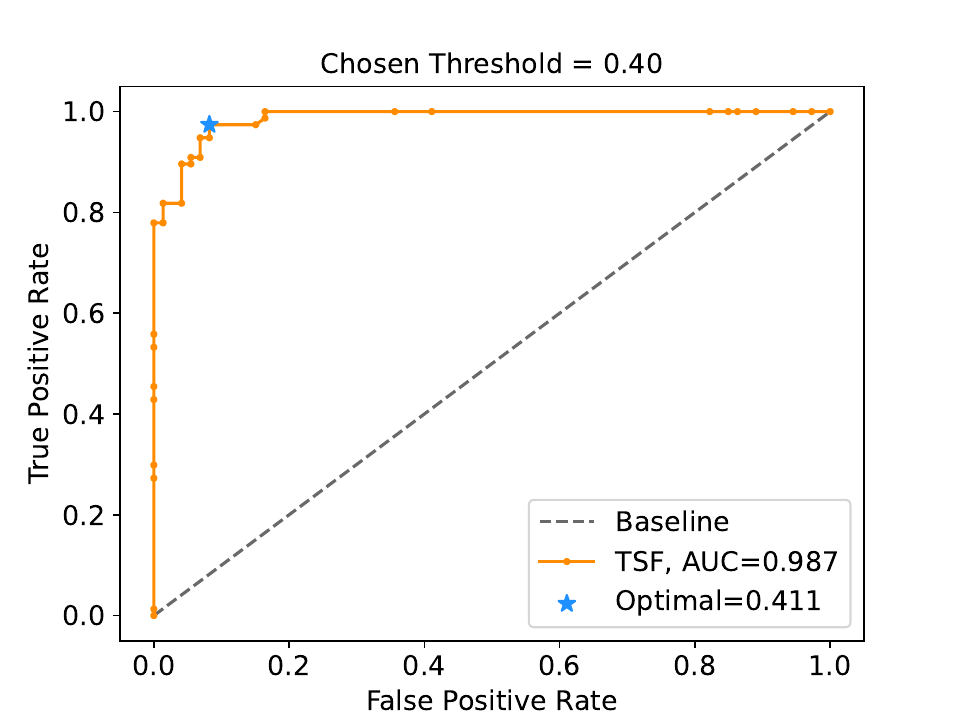}{0.49\textwidth}{(a)}
          \fig{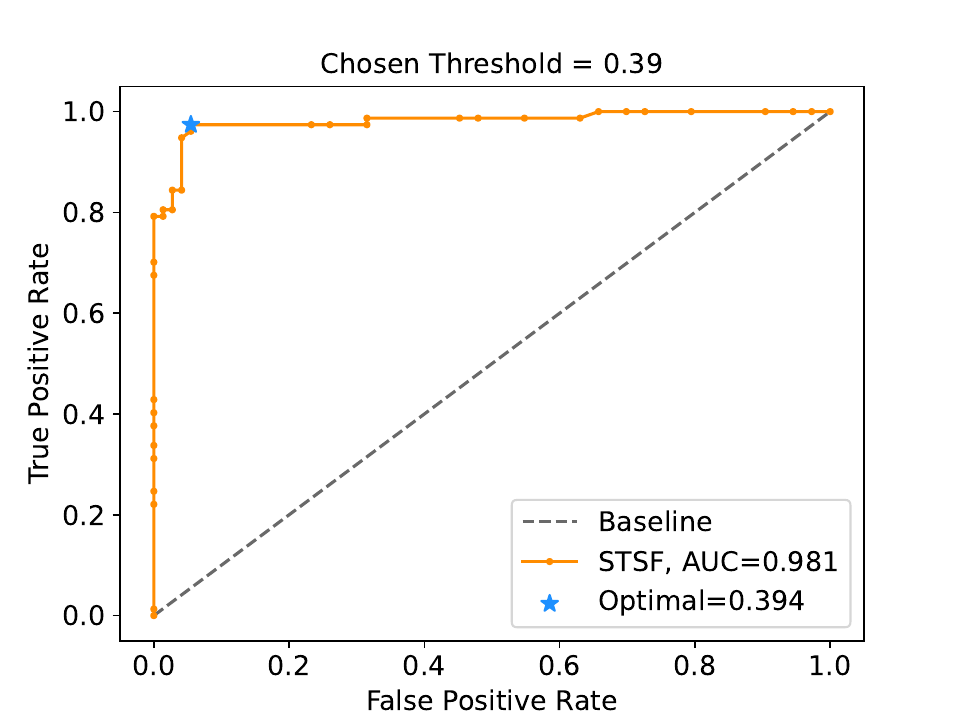}{0.49\textwidth}{(b)}
          }
\gridline{\fig{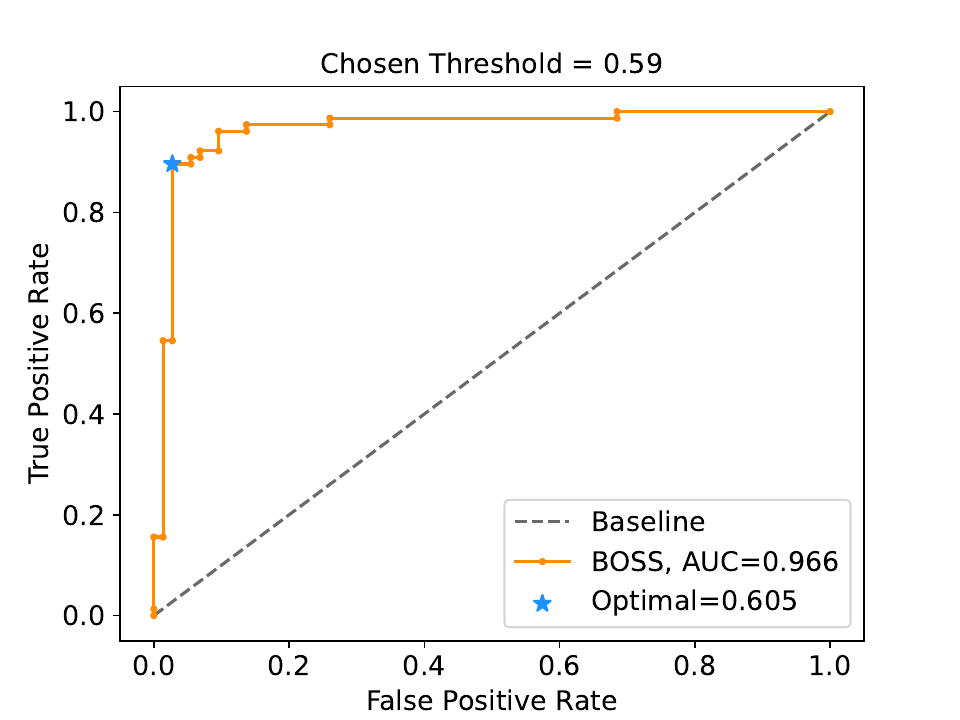}{0.49\textwidth}{(c)}
          }
\caption{Receiver Operating Characteristic (ROC) curves for (a) TSF, (b) STSF and (c) BOSS models on the test set with the area under the curve (AUC) inset in the legend. Here, the x-axis shows the false positive rate (FPR) and the y-axis shows the true positive rate (TPR) for the classifier. The dashed diagonal line indicates the ROC curve for a baseline or no-skill classifier. A starred point in blue color positioned at the top left of the plot indicates the optimal threshold value of the model. In addition, the chosen threshold to estimate the model skills is provided at the top of the plot for the model, respectively.}
\label{fig:roc}
\end{figure*}

In Figure \ref{fig:roc}, we show the ROC curves for our models based on the TPR and FPR. Here, we indicate the optimal threshold of the classifiers in the upper-left corner of the ROC curve (as a blue star). Furthermore, the TSF has an ROC-AUC of 0.987, STSF  has 0.981, and for BOSS, we get 0.966, indicating excellent discriminatory performance in all the classifiers. The skill scores and model evaluation discussed further are based on the specific chosen (that gives optimal results) threshold after our initial analysis: TSF = 0.40 (Fig. \ref{fig:roc}a), STSF = 0.39 (Fig. \ref{fig:roc}b), and BOSS = 0.59 (Fig. \ref{fig:roc}c). In Appendix \ref{sec:TA}, we provide an evaluation of the influence of varying thresholds on the scores as shown in Figure \ref{fig:comp}.

A 2 $\times$ 2 contingency table constitutes the following elements: true positives (TP), true negatives (TN), false positives (FP), and false negatives (FN). Here, TP indicates the number of correctly predicted large SEP events (positive class) by a model, while TN represents the number of correctly predicted small SEP events (negative class). FP corresponds to the number of small events predicted as large (false alarms), while FN corresponds to the number of large events predicted as small (misses). Subsequently, the aim of our best model should be to reduce incorrect results represented by both FP and FN. In Table \ref{tab:cm}, we show the contingency tables based on the chosen classification threshold of our models on the test set. TSF and STSF indicate a relatively higher number of false alarms, but the BOSS model outputs a fairly close number of misses and false alarms.

\begin{deluxetable*}{ll|c|c|c|c|c|c|}[ht!]
\tablenum{3}
\tablecaption{Contingency tables for the models on the test set. \label{tab:cm}}
\tablewidth{0pt}
\tablehead{
\textbf{}   &\textbf{}   &  \multicolumn{2}{c|}{\textbf{TSF}} & \multicolumn{2}{c|}{\textbf{STSF}} & \multicolumn{2}{c|}{\textbf{BOSS}}   \\
\colhead{} & \colhead{} & \colhead{} & \colhead{} & \colhead{} & \colhead{} & \colhead{} & \colhead{} \\
\textbf{}   &  \textbf{}   &  \multicolumn{2}{c|}{Predicted} & \multicolumn{2}{c|}{Predicted} & \multicolumn{2}{c|}{Predicted}   \\
 &  & Large & Small & Large & Small & Large & Small \\
}
\startdata
 & Large & 75 & 2 & 75 & 2 & 71 & 6 \\
True &||| &|||&|||& ||| & ||| & ||| & ||| \\
 & Small & 6 & 67 & 4  & 69 & 5 & 68\\
\enddata
\tablecomments{Truth tables based on the chosen classification threshold for all the models on the test set. The first column is a shared entry of true labels against predictive labels for each corresponding model. The elements indicate the number of predictions with respect to the actual occurrences in the test set.
\newline
Model names: TSF - Time Series Forest; STSF - Supervised Time Series Forest; BOSS - Bag of SFA Symbols}
\end{deluxetable*}

Focusing on the importance of positive classes, we consider the F$_{1}$-score defined in Equation \ref{sc:f1_wt}. It ranges between 0 and 1 such that scores closer to 1 indicate the model to be better. To account for the FPR, that is, compare the difference between the probability of detection and the probability of false detection, we utilize true skill statistics (TSS; Woodcock \citeyear{woodcock_evaluation_1976}; Dann \citeyear{daan1985forecast}) as shown in Equation \ref{sc:tss}. TSS ranges from -1 to +1, where the latter indicates a perfect score. TSS $\leq$0 indicates agreement no better than a random classification.

\begin{equation} \label{sc:tss}
TSS = \frac{(TP \times TN) - (FP \times FN)}{(TP + FN) \times (FP + TN)}
\end{equation}

Furthermore, the Heidke skill score (HSS; Heidke \citeyear{heidke_berechnung_1926}) measures the improvement of the forecast over a random prediction as defined in Equation \ref{sc:hss}. HSS with 1 indicates perfect performance and 0 indicates no skill. A no-skill means the forecast is not better than a random binary forecast based on class distributions.

\begin{equation} \label{sc:hss}
HSS = \frac{2 \times ((TP \times TN) - (FP \times FN))}{((TP + FN) \times (TN + FN)) + ((FP + TN) \times (FP + TP))}
\end{equation}

The Gilbert Skill Score (GSS; Schaefer \citeyear{schaefer1990critical}) considers the number of hits due to chance, which is the frequency of an event multiplied by the total number of forecast events. This score formula is given by Equation \ref{sc:gss}. GSS ranges from -1/3 to 1. Here, 0 indicates no skill, while 1 is a perfect forecast.

\begin{equation} \label{sc:gss}
GSS = \frac{TP - (\frac{(TP + FN) \times (TP + FP)}{TP + FP + TN + FN})}{(TP + FP + FN) - (\frac{(TP + FN) \times (TP + FP)}{TP + FP + TN + FN})}
\end{equation}

However, accounting for the true negatives to assess the performance of a binary class problem is essential in our context. Hence, we also choose Matthew's correlation coefficient (MCC) as defined in Equation \ref{sc:mcc}. MCC ranges from -1 to 1. Here, 0 indicates no skill, while 1 shows perfect agreement predicted and actual values.

\begin{equation} \label{sc:mcc}
MCC = \frac{(TP \times TN) - (FP \times FN)}{\sqrt{(TP + FP) \times (TP + FN) \times (TP + FP) \times (TN + FN)}}
\end{equation}

\begin{deluxetable*}{lccccc}[ht!]
\tablenum{4}
\tablecaption{Model performances on the test set. \label{tab:metrics}}
\tablewidth{0pt}
\tablehead{
\textbf{Model} & \textbf{F$_{1}$} & \textbf{TSS} & \textbf{HSS} & \textbf{GSS} & \textbf{MCC}\\
}
\startdata
TSF   & 0.947 & 0.892 & 0.893  &  0.807  &  0.894\\
STSF  & \textbf{0.960} & \textbf{0.919} & \textbf{0.920} & \textbf{0.852} & \textbf{0.920}\\
BOSS  & 0.927 & 0.854 & 0.8533  &  0.744  &  0.853\\
\enddata
\tablecomments{Class metrics are presented here for the best models implemented as an ensemble of univariate classifiers on the test set. \newline
Model names: TSF - Time Series Forest; STSF - Supervised Time Series Forest; BOSS - Bag of SFA Symbols \newline
Metric names: TSS - True skill statistics; HSS - Heidke skill score; GSS - Gilbert skill score; MCC - Matthews correlation coefficient.
}
\end{deluxetable*}
We approach the SEP event prediction problem from a time series classification perspective using the GSEP data set. The skill scores based on the respective chosen classification threshold for all our classifiers on the test set are presented in Table \ref{tab:metrics}. One can see that the STSF model performs well compared to the TSF and BOSS models in terms of all the scores.

As there is no one-to-one correspondence between the task, data set, and sampling implemented, we do not extensively compare our results with earlier studies. In Table \ref{tab:seps}, we list existing models that implement empirical or ML methods for predicting E$\geq$10 MeV SEP events. The models in these studies have been developed focusing on a combination of various solar parameters, including solar flare X-ray fluxes and their properties. As can be seen, the period considered in these studies varies depending on the availability of their desired data set. We include two common metrics; HSS and TSS (where available) used across these works in the table. HSS is an advanced metric and is highly dependent on the number of samples present in each binary class of a data set (Bobra \& Couvidat \citeyear{bobra2015solar}).

While we make short-term predictions, other works typically focus on forecasting SEP event onset hours and days ahead. Moreover, no previous work has focused on the classification task between large and small SEP events. Nonetheless, in addition to other evaluation methods demonstrated in this paper, our results show great performance potential in using column ensembles of the time series ML. The interval-based STSF model architecture demonstrated in this paper promises to be helpful to be implemented in NRT operations. In Appendix \ref{sec:ER}, we show the effect of randomness in the TSF and STSF architectures on the optimal threshold for classification and further establish confidence and robustness in our predictions. Therefore, our future work will transform the capacity of the STSF model to provide short-term predictions on NRT data.

\begin{deluxetable*}{lcccc}[ht!]
\tablenum{5}
\tablecaption{List of existing SEP event prediction models that consider solar protons, X-ray flare fluxes, and their properties as input. \label{tab:seps}}
\tablewidth{0pt}
\tablehead{
\textbf{Model}  & \textbf{Period}  & \textbf{Type} & \textbf{HSS} & \textbf{TSS}}
\startdata
Balch (\citeyear{balch}) & 1986 - 2004 & Empirical & 0.48$\pm$0.04 & - \\
Laurenza et al. (\citeyear{laurenza2009technique}) & 1995 - 2005 & Empirical & 0.58 &-\\
Winter \& Ledbetter (\citeyear{winter2015ApJ}) & 1995 - 2005 & Empirical & 0.60 &-\\
Alberti et al. (\citeyear{alberti2017solar}) & 2004 - 2014 & Empirical & 0.55  & - \\
Anastasiadis et al. (\citeyear{anastasiadis2017predicting}) & 1984 - 2013 & Empirical & 0.37$\pm$0.011  & 0.5\\
Engell et al. (\citeyear{engell2017sprints}) & 1986 - 2018 & ML & 0.58 & -\\
Papaioannou et al. (\citeyear{papaioannou2018nowcasting}) & 1997 - 2013 & Empirical & 0.65 & -\\
Lavasa et al. (\citeyear{lavasa2021assessing}) & 1988 - 2013 & ML & 0.69$\pm$0.04  & 0.75$\pm$0.05 \\
Aminalragia-Giamini et al. (\citeyear{aminalragia2021solar}) & 1988 - 2013 & ML & - & 0.79\\
Sadykov et al. (\citeyear{sadykov2021prediction}) & 2010 - 2019 & ML & 0.434$\pm$0.046 & 0.821$\pm$0.003\\
\enddata
\tablecomments{HSS - Heidke skill score; TSS - true skill statistics; ML - Machine Learning}
\end{deluxetable*}

\section{Conclusions} \label{sec:conclusion}
Solar energetic particle (SEP) events are one of the main elements of space weather, along with solar flares and coronal mass ejections. Towards predictive efforts of SEP events, we utilize the recently developed GSEP data set (Rotti et al. \citeyear{rotti2022}) publicly available from Harvard Dataverse \dataset[10.7910/DVN/DZYLHK]{https://doi.org/10.7910/DVN/DZYLHK}. The data set constitutes \textit{in situ} time series measurements from the NOAA-GOES missions for solar cycles 22 to 24. They are long band (1–8Å) X-ray measurements from the XRS instrument and proton fluxes (P3, P5, P7) from the SEM instrument. We use these parameters to evaluate the performance of our multivariate time series models.

The target labels are defined based on integral proton fluxes ($I_{P}$) recorded by the GOES P3 channel. Positive labels are large SEP events crossing the 10 pfu threshold; negative otherwise. There are 433 SEP events in the GSEP data set, of which 244 are large. We consider a fixed length of 12 hours minus five minutes of fluxes before the SEP event onset constitutes the observation window. Therefore, the total length for each time series corresponds to 715 instances.

Our focus in the present work is to see whether the model can classify the P3 proton channel flux to be crossing the 10 pfu limit or not. In other words, if the 10 pfu limit is outset in the 10 MeV channel, then the model outputs a ``true'' or ``yes'' label indicating a large event. If not, then it is a small or sub-event. When implemented in NRT operation, the yes/no outputs from the models are in succession for the next few minutes of the prediction window.

Machine learning (ML) methods are at the forefront of the latest techniques in space weather forecasting. The crucial focus on implementing ML towards SEP event forecasting is for the upcoming NASA human missions to the Moon and Mars (Whitman et al. \citeyear{WHITMAN2022}). In this scenario, short-term forecasts become relevant and require distinct attention to precise and sensitive prediction of large SEP event occurrences. This work implements time series-based ML models in a binary classification schema. Because no single algorithm always creates the best results, we want to experiment with multiple models and evaluate their performances.

Interval-based methods are based on splitting the time series into phase-dependent distinct intervals. Statistics are gathered from each interval to fit individual classifiers on the data. The final classification is assigned based on majority voting of the most common class generated by the individual classifiers. We consider two interval-based classifiers in our work. They are time series forest (TSF) and supervised time series forest (STSF). TSF is a collection of decision trees applied to the feature sets (mean, standard deviation and slope) extracted from the intervals. Here, the average prediction from each tree is obtained, and based on a majority vote, the final output is predicted. STSF builds on the TSF model by implementing a metric to supervise the random sampling such that the subsamples represent the entire series. Statistical features such as mean, median, standard deviation, slope, min, max and interquartile range are extracted from each interval for three representations (time, frequency and derivative). The classifier then concatenates these extracted values to form a new dataset and builds a random forest model to make predictions. Another model we implement is the BOSS ensemble, a dictionary-based algorithm. In that, small intervals of length `l' are transformed into ``words" and stored as histograms for each input time series. The occurrence of the word during prediction is used to classify the series to a label on a weighted output. 

The learning curves of our classifiers indicate sufficient data used during the training phase. On the test set, we estimate the confidence intervals of the predictions using reliability diagrams and use Brier score loss in our evaluation strategy. We construct the ROC curve for our models and identify the best classification threshold to transform the probabilistic decisions into binary labels. We use the area under ROC curve (AUC), F$_{1}$-score, true skill statistics (TSS), Gilbert skill score (GSS), Heidke skill score (HSS), and Matthews correlation coefficient (MCC) to further assess the performances of our models.

The results in this paper shows that the STSF classifier performs well compared to the TSF and BOSS models. Multiple evaluation schemes relatively indicate that our model obtains the best scores compared to existing methods but in the framework of SEP event classification. In addition, our work shows that interval-based classifiers have great potential to improve short-term forecasts, and an ensemble model is a suitable predictor for use in an operational context.

The SEP prediction model we have developed in this paper is very high confidence. Our objective is to develop a short-term SEP event forecasting algorithm to predict whether the solar proton flux level will surpass the SWPC `S1' threshold. In that respect, our approach is very different from the standard SEP prediction methods, which forecast the likelihood of an SEP storm in the coming 24 or 48 hours. Our model would allow for SEP warnings to be called off at the last minute and for high-level (E$\geq$10 MeV) SEP event forecasts to be confirmed with high certainty or issued if there is no longer-term alert. Certainly, the latter case will be extremely valuable for Artemis astronauts in extra-vehicular activities (EVAs) or on the surface of the Moon. If reliable, our model will give the real-time forecasters at the Space Radiation Analysis Group (SRAG) a useful tool to help them decide whether to issue an alert. In an operational setting, we envisage our system to sit on top of forecasts with a much longer prediction horizon but lower precision, such as current forecasts\footnote{\url{https://ccmc.gsfc.nasa.gov/scoreboards/}}.

More avenues can be explored for future work, which includes but is not limited to extending the analysis to (1) consider ``no-SEP" phases i.e., SEP-quite periods following the occurrence of large ($\geq$M1.0) flares, and (2) build different ensemble strategies.

\section{Acknowledgments}
We thank the anonymous reviewer for constructive comments on the manuscript that have improved the contents of the paper. We acknowledge the use of X-ray and proton flux data from the GOES missions made available by NOAA. Authors Petrus Martens' and Berkay Aydin's contribution to this work is supported by NASA SWR2O2R Grant 80NSSC22K0272. Author Sumanth Rotti carried out this work while supported by the NASA FINESST Grant 80NSSC21K1388.


\software{pandas (McKinnet et al. \citeyear{mckinney2010data}), numpy (Van Der Walt et al. \citeyear{van2011numpy}; Harris et al. \citeyear{harris2020array}), sklearn (Pedregosa et al. \citeyear{scikit-learn}), sktime (L{\"o}ning et al. \citeyear{loning2019sktime, markussktime}), matplotlib (Hunter \citeyear{hunter2007matplotlib})}

\appendix
\section{Threshold Analysis}\label{sec:TA}

The classification threshold is the decision threshold that allows us to map the probabilistic output of a classifier to a binary category. In other words, it is a cut-off point used to assign a specific predicted class label for each sample. In our model analysis phase, we used the Receiver Operating Characteristic (ROC) curve, which is a diagnostic tool used to evaluate a set of probabilistic predictions made by a model. The ROC curve is useful for understanding the trade-off between true positive rate (TPR) and false positive rate (FPR) at different thresholds.

By default, the classification threshold in our models is 0.5. Any prediction above 0.5 belongs to the positive class, and that below 0.5 belongs to the negative class. However, 0.5 is not always optimal, and we identify a reliable threshold for the classifier that better splits between the two target classes. That is, we choose the threshold that provides a TPR with an acceptable FPR to make decisions using the classifier.

In the present work, we find the optimal threshold using the Youden Index (J; Youden \citeyear{youden1950index}) defined in Equation \ref{sc:J}. Here, sensitivity is TPR and specificity is (1-FPR). Therefore, by estimating TPR-FPR for each threshold, we obtain a maximum J as a cut-off point that optimizes classification between the two classes. The obtained best J-value gives us the optimal threshold of the classifier.

\begin{figure*}[ht!]
\gridline{\fig{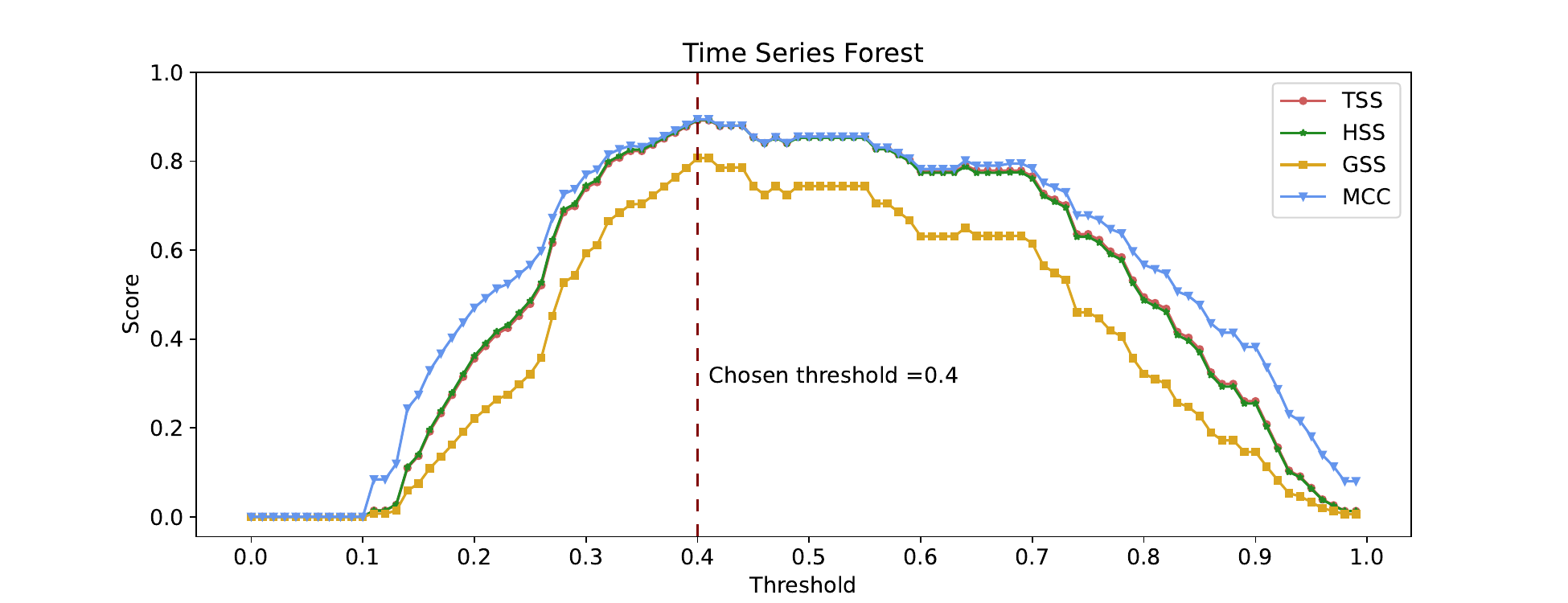}{0.85\textwidth}{(a)}
          }
\gridline{\fig{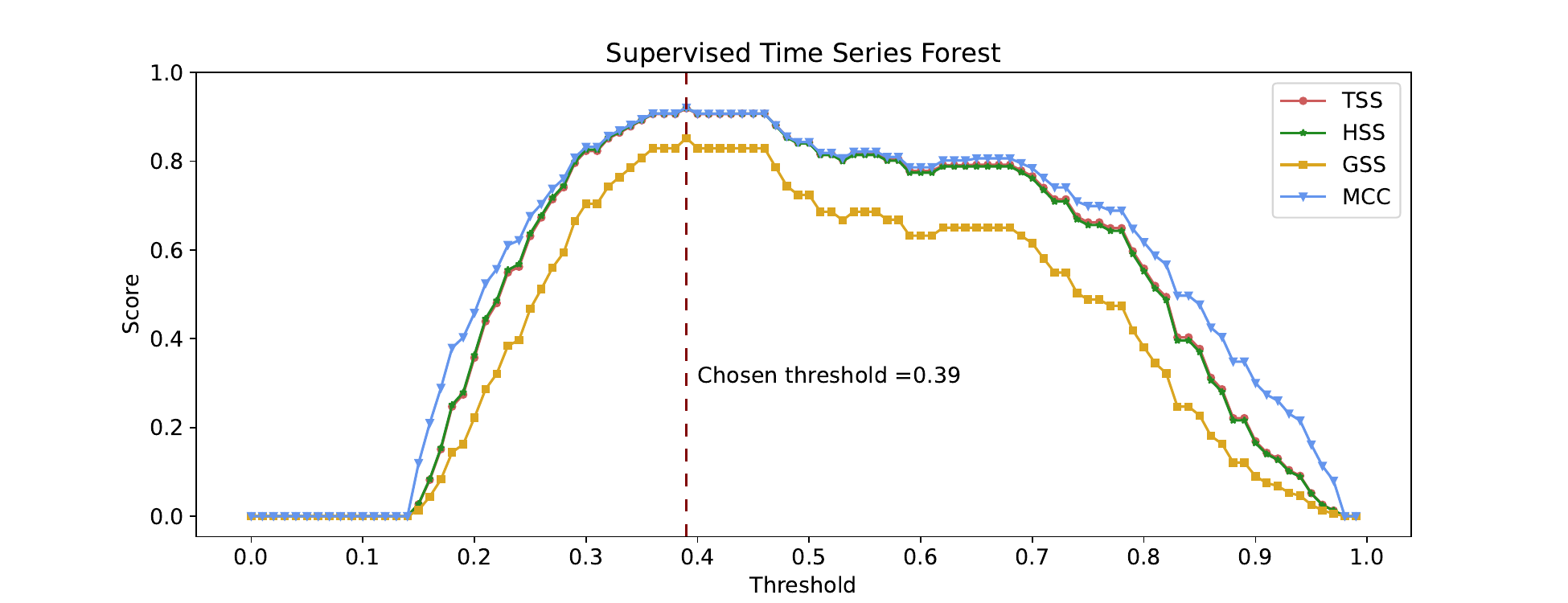}{0.85\textwidth}{(b)}
          }
\gridline{\fig{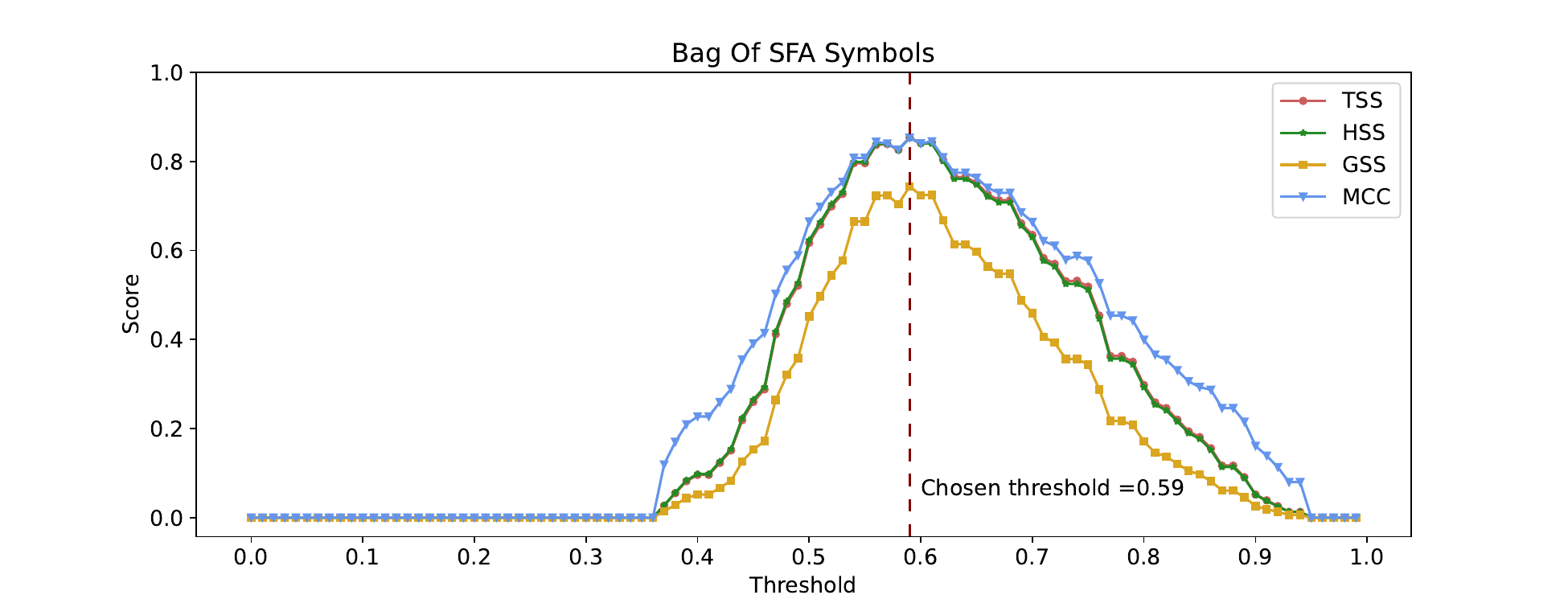}{0.85\textwidth}{(c)}
          }
\caption{Variation in skills such as TSS, HSS, GSS and MCC with respect to increasing the classification threshold for (a) TSF, (b) STSF, and (c) BOSS models on the test set. The optimal threshold value for each model is inset in the plot.
\label{fig:comp}}
\end{figure*}

Furthermore, we demonstrate the effect of ``thresholding'' on the model performances by visualizing the variations in the skills due to changing thresholds. For this purpose, we used advanced metrics discussed in Section \ref{sec:results}. We define a set of thresholds (from 0.0 to 1.0) and then evaluate predicted probabilities under each threshold. That is, we transform/binarize the predicted probabilities into labels for the respective threshold and estimate the skill scores in order to find and select the best threshold value. Figure \ref{fig:comp} shows the influence of variation in the classification threshold for each model. The TSF (Fig. \ref{fig:comp}a) and STSF (Fig. \ref{fig:comp}b) have a very close optimal threshold that is less than 50\%. The BOSS model (Fig. \ref{fig:comp}c) shows optimal performance at a threshold of $\approx$ 60\%.

\section{Effect of Randomness}\label{sec:ER}
\begin{figure}[ht!]
\plotone{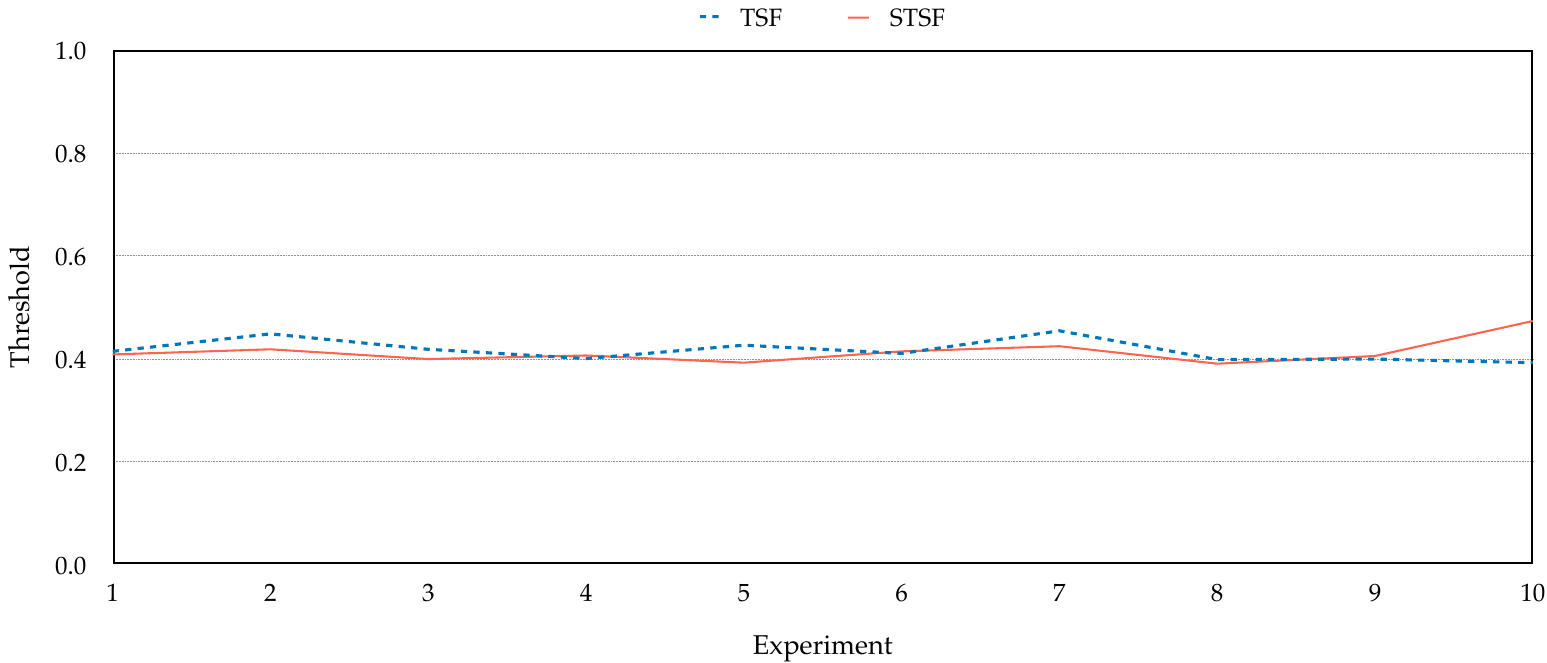}
\caption{Experimantal evaluation of the impact of random components in the TSF and STSF model structures on optimal classification threshold. Here, the y-axis shows the thresholds (in the range of 0.0 to 1.0) and the x-axis shows the number of experiments. The median (mean) threshold for TSF is 0.412 (0.415) and for STSF it is 0.407 (0.412).
\label{fig:rand}}
\end{figure}

Of the three models considered in this work, TSF considers random intervals from the input time series and implements a random forest to fit the feature vectors and make predictions. Although STSF largely overcomes the randomization of interval selection, it consists of a tree-based random forest structure at its core. Because TSF and STSF models have random components in their architecture, we run both models multiple (10) times and find the variations in their respective optimal threshold values as shown in Figure \ref{fig:rand}. The median (mean) value for TSF is 0.412 (0.415), and for STSF it is 0.407 (0.412). Comparing the above values with the chosen thresholds (as shown in Figure \ref{fig:comp}) for the respective classifiers, we are confident of our model predictions and their capabilities to be further transformed for operational standards.

\bibliography{sample631}{}

\begin{thebibliography}{}
\expandafter\ifx\csname natexlab\endcsname\relax\def\natexlab#1{#1}\fi
\providecommand{\url}[1]{\href{#1}{#1}}
\providecommand{\dodoi}[1]{doi:~\href{http://doi.org/#1}{\nolinkurl{#1}}}
\providecommand{\doeprint}[1]{\href{http://ascl.net/#1}{\nolinkurl{http://ascl.net/#1}}}
\providecommand{\doarXiv}[1]{\href{https://arxiv.org/abs/#1}{\nolinkurl{https://arxiv.org/abs/#1}}}

\bibitem[{Ahmadzadeh {et~al.}(2019)Ahmadzadeh, Hostetter, Aydin, Georgoulis, Kempton, Mahajan, \& Angryk}]{ahmadzadeh2019challenges}
Ahmadzadeh, A., Hostetter, M., Aydin, B., {et~al.} 2019, in 2019 IEEE international conference on big data (Big Data), Ieee, 1423--1431

\bibitem[{Alberti {et~al.}(2017)Alberti, Laurenza, Cliver, Storini, Consolini, \& Lepreti}]{alberti2017solar}
Alberti, T., Laurenza, M., Cliver, E., {et~al.} 2017, The Astrophysical Journal, 838, 59

\bibitem[{Aminalragia-Giamini {et~al.}(2021)Aminalragia-Giamini, Raptis, Anastasiadis, Tsigkanos, Sandberg, Papaioannou, Papadimitriou, Jiggens, Aran, \& Daglis}]{aminalragia2021solar}
Aminalragia-Giamini, S., Raptis, S., Anastasiadis, A., {et~al.} 2021, Journal of Space Weather and Space Climate, 11, 59

\bibitem[{Anastasiadis {et~al.}(2017)Anastasiadis, Papaioannou, Sandberg, Georgoulis, Tziotziou, Kouloumvakos, \& Jiggens}]{anastasiadis2017predicting}
Anastasiadis, A., Papaioannou, A., Sandberg, I., {et~al.} 2017, Solar Physics, 292, 1

\bibitem[{Arbib(2003)}]{arbib2003handbook}
Arbib, M.~A. 2003, The handbook of brain theory and neural networks (MIT press)

\bibitem[{Bagnall {et~al.}(2017)Bagnall, Lines, Bostrom, Large, \& Keogh}]{bagnall2017great}
Bagnall, A., Lines, J., Bostrom, A., Large, J., \& Keogh, E. 2017, Data mining and knowledge discovery, 31, 606

\bibitem[{Bain {et~al.}(2021)Bain, Steenburgh, Onsager, \& Stitely}]{bain2021summary}
Bain, H., Steenburgh, R., Onsager, T., \& Stitely, E. 2021, Space Weather, 19, e2020SW002670

\bibitem[{{Balch}(2008)}]{balch}
{Balch}, C.~C. 2008, Space Weather, 6, S01001, \dodoi{10.1029/2007SW000337}

\bibitem[{Beck {et~al.}(2005)Beck, Latocha, Rollet, \& Stehno}]{beck2005tepc}
Beck, P., Latocha, M., Rollet, S., \& Stehno, G. 2005, Advances in Space Research, 36, 1627

\bibitem[{Bobra \& Couvidat(2015)}]{bobra2015solar}
Bobra, M.~G., \& Couvidat, S. 2015, The Astrophysical Journal, 798, 135

\bibitem[{Bornmann {et~al.}(1996)Bornmann, Speich, Hirman, Matheson, Grubb, Garcia, \& Viereck}]{bornman}
Bornmann, P.~L., Speich, D., Hirman, J., {et~al.} 1996, in GOES-8 and Beyond, ed. E.~R. Washwell, Vol. 2812, International Society for Optics and Photonics (SPIE), 291 -- 298, \dodoi{10.1117/12.254076}

\bibitem[{Boubrahimi {et~al.}(2017)Boubrahimi, Aydin, Martens, \& Angryk}]{boubrahimi2017prediction}
Boubrahimi, S.~F., Aydin, B., Martens, P., \& Angryk, R. 2017, in 2017 IEEE International Conference on Big Data (Big Data), IEEE, 2533--2542

\bibitem[{Cabello {et~al.}(2020)Cabello, Naghizade, Qi, \& Kulik}]{cabello2020fast}
Cabello, N., Naghizade, E., Qi, J., \& Kulik, L. 2020, in 2020 IEEE International Conference on Data Mining (ICDM), IEEE, 948--953

\bibitem[{{Camporeale}(2019)}]{enrico2019}
{Camporeale}, E. 2019, Space Weather, 17, 1166, \dodoi{10.1029/2018SW002061}

\bibitem[{Cane(1995)}]{CANE199535}
Cane, H. 1995, Nuclear Physics B - Proceedings Supplements, 39, 35, \dodoi{https://doi.org/10.1016/0920-5632(95)00005-T}

\bibitem[{{Cane} {et~al.}(1986){Cane}, {McGuire}, \& {von Rosenvinge}}]{1986cane}
{Cane}, H.~V., {McGuire}, R.~E., \& {von Rosenvinge}, T.~T. 1986, \apj, 301, 448, \dodoi{10.1086/163913}

\bibitem[{Cassisi {et~al.}(2012)Cassisi, Montalto, Aliotta, Cannata, Pulvirenti, {et~al.}}]{cassisi2012similarity}
Cassisi, C., Montalto, P., Aliotta, M., {et~al.} 2012, Advances in data mining knowledge discovery and applications, 71

\bibitem[{Cliver \& D’Huys(2018)}]{cliver2018size}
Cliver, E.~W., \& D’Huys, E. 2018, The Astrophysical Journal, 864, 48

\bibitem[{Daan(1985)}]{daan1985forecast}
Daan, H. 1985, Statistics, and Decision Making in the Atmospheric Sciences, AH Murphy and RW Katz, Eds., Westview Press, 379

\bibitem[{Deng {et~al.}(2013)Deng, Runger, Tuv, \& Vladimir}]{deng2013time}
Deng, H., Runger, G., Tuv, E., \& Vladimir, M. 2013, Information Sciences, 239, 142

\bibitem[{Dierckxsens {et~al.}(2015)Dierckxsens, Tziotziou, Dalla, Patsou, Marsh, Crosby, Malandraki, \& Tsiropoula}]{dierckxsens2015relationship}
Dierckxsens, M., Tziotziou, K., Dalla, S., {et~al.} 2015, Solar Physics, 290, 841

\bibitem[{Engell {et~al.}(2017)Engell, Falconer, Schuh, Loomis, \& Bissett}]{engell2017sprints}
Engell, A., Falconer, D., Schuh, M., Loomis, J., \& Bissett, D. 2017, Space Weather, 15, 1321

\bibitem[{Falconer {et~al.}(2011)Falconer, Barghouty, Khazanov, \& Moore}]{falconer2011tool}
Falconer, D., Barghouty, A.~F., Khazanov, I., \& Moore, R. 2011, Space Weather, 9

\bibitem[{Faouzi(2022)}]{faouzi2022time}
Faouzi, J. 2022, Machine Learning (Emerging Trends and Applications)

\bibitem[{Fulcher \& Jones(2014)}]{fulcher2014highly}
Fulcher, B.~D., \& Jones, N.~S. 2014, IEEE Transactions on Knowledge and Data Engineering, 26, 3026

\bibitem[{Gopalswamy {et~al.}(2001)Gopalswamy, Lara, Yashiro, Kaiser, \& Howard}]{gopalswamy2001predicting}
Gopalswamy, N., Lara, A., Yashiro, S., Kaiser, M.~L., \& Howard, R.~A. 2001, Journal of Geophysical Research: Space Physics, 106, 29207

\bibitem[{{Gopalswamy} {et~al.}(2017){Gopalswamy}, {M{\"a}kel{\"a}}, {Yashiro}, {Thakur}, {Akiyama}, \& {Xie}}]{gopalswamy2017hierarchical}
{Gopalswamy}, N., {M{\"a}kel{\"a}}, P., {Yashiro}, S., {et~al.} 2017, in Journal of Physics Conference Series, Vol. 900, Journal of Physics Conference Series, 012009, \dodoi{10.1088/1742-6596/900/1/012009}

\bibitem[{Gopalswamy {et~al.}(2008)Gopalswamy, Yashiro, Xie, Akiyama, Aguilar-Rodriguez, Kaiser, Howard, \& Bougeret}]{gopalswamy2008radio}
Gopalswamy, N., Yashiro, S., Xie, H., {et~al.} 2008, The Astrophysical Journal, 674, 560

\bibitem[{{Grubb}(1975)}]{grubb}
{Grubb}, R.~N. 1975, {The SMS/GOES space environment monitor subsystem}, NASA STI/Recon Technical Report N

\bibitem[{Hansen \& Salamon(1990)}]{hansen1990neural}
Hansen, L.~K., \& Salamon, P. 1990, IEEE transactions on pattern analysis and machine intelligence, 12, 993

\bibitem[{Harris {et~al.}(2020)Harris, Millman, Van Der~Walt, Gommers, Virtanen, Cournapeau, Wieser, Taylor, Berg, Smith, {et~al.}}]{harris2020array}
Harris, C.~R., Millman, K.~J., Van Der~Walt, S.~J., {et~al.} 2020, Nature, 585, 357

\bibitem[{Hastie {et~al.}(2009)Hastie, Tibshirani, Friedman, \& Friedman}]{hastie2009elements}
Hastie, T., Tibshirani, R., Friedman, J.~H., \& Friedman, J.~H. 2009, The elements of statistical learning: data mining, inference, and prediction, Vol.~2 (Springer)

\bibitem[{Heidke(1926)}]{heidke_berechnung_1926}
Heidke, P. 1926, Geografiska Annaler, 8, 301, \dodoi{10.1080/20014422.1926.11881138}

\bibitem[{Hunter(2007)}]{hunter2007matplotlib}
Hunter, J.~D. 2007, Computing in science \& engineering, 9, 90

\bibitem[{Jackman \& McPeters(1987)}]{jackman1987solar}
Jackman, C.~H., \& McPeters, R.~D. 1987, Physica Scripta, 1987, 309

\bibitem[{Ji {et~al.}(2021)Ji, Arya, Kempton, Angryk, Georgoulis, \& Aydin}]{Ji2021}
Ji, A., Arya, A., Kempton, D., {et~al.} 2021, in 2021 {IEEE} Third International Conference on Cognitive Machine Intelligence ({CogMI}) ({IEEE}), \dodoi{10.1109/cogmi52975.2021.00022}

\bibitem[{Ji {et~al.}(2020)Ji, Aydin, Georgoulis, \& Angryk}]{ji2020all}
Ji, A., Aydin, B., Georgoulis, M.~K., \& Angryk, R. 2020, in 2020 IEEE International Conference on Big Data (Big Data), IEEE, 4218--4225

\bibitem[{Jiggens {et~al.}(2019)Jiggens, Clavie, Evans, O'Brien, Witasse, Mishev, Nieminen, Daly, Kalegaev, Vlasova, {et~al.}}]{jiggens2019situ}
Jiggens, P., Clavie, C., Evans, H., {et~al.} 2019, Space Weather, 17, 99

\bibitem[{{Kahler}(1992)}]{1992kahler}
{Kahler}, S.~W. 1992, \araa, 30, 113, \dodoi{10.1146/annurev.aa.30.090192.000553}

\bibitem[{{Kahler} {et~al.}(2007){Kahler}, {Cliver}, \& {Ling}}]{2007kahler}
{Kahler}, S.~W., {Cliver}, E.~W., \& {Ling}, A.~G. 2007, Journal of Atmospheric and Solar-Terrestrial Physics, 69, 43, \dodoi{10.1016/j.jastp.2006.06.009}

\bibitem[{Keogh {et~al.}(2001)Keogh, Chakrabarti, Pazzani, \& Mehrotra}]{keogh2001dimensionality}
Keogh, E., Chakrabarti, K., Pazzani, M., \& Mehrotra, S. 2001, Knowledge and information Systems, 3, 263

\bibitem[{Laurenza {et~al.}(2009)Laurenza, Cliver, Hewitt, Storini, Ling, Balch, \& Kaiser}]{laurenza2009technique}
Laurenza, M., Cliver, E., Hewitt, J., {et~al.} 2009, Space Weather, 7

\bibitem[{Lavasa {et~al.}(2021)Lavasa, Giannopoulos, Papaioannou, Anastasiadis, Daglis, Aran, Pacheco, \& Sanahuja}]{lavasa2021assessing}
Lavasa, E., Giannopoulos, G., Papaioannou, A., {et~al.} 2021, Solar Physics, 296, 107

\bibitem[{L{\"o}ning {et~al.}(2019)L{\"o}ning, Bagnall, Ganesh, Kazakov, Lines, \& Kir{\'a}ly}]{loning2019sktime}
L{\"o}ning, M., Bagnall, A., Ganesh, S., {et~al.} 2019, arXiv preprint, \dodoi{10.48550/arXiv.1909.07872}

\bibitem[{Löning {et~al.}(2022)Löning, Király, Bagnall, Middlehurst, Ganesh, Oastler, Lines, Walter, ViktorKaz, Mentel, \& et~al.}]{markussktime}
Löning, M., Király, F., Bagnall, T., {et~al.} 2022, sktime/sktime: v0.13.4,  Zenodo, \dodoi{10.5281/zenodo.7117735}

\bibitem[{Manning {et~al.}(2008)Manning, Raghavan, \& Schütze}]{manning_raghavan_schütze_2008}
Manning, C.~D., Raghavan, P., \& Schütze, H. 2008, Introduction to Information Retrieval (Cambridge University Press), \dodoi{10.1017/CBO9780511809071}

\bibitem[{Marqu{\'e} {et~al.}(2006)Marqu{\'e}, Posner, \& Klein}]{marque2006solar}
Marqu{\'e}, C., Posner, A., \& Klein, K.-L. 2006, The Astrophysical Journal, 642, 1222

\bibitem[{McKinney {et~al.}(2010)}]{mckinney2010data}
McKinney, W., {et~al.} 2010, Data structures for statistical computing in python

\bibitem[{Murphy(1973)}]{murphy1973new}
Murphy, A.~H. 1973, Journal of Applied Meteorology and Climatology, 12, 595

\bibitem[{{N{\'u}{\~n}ez}(2011)}]{2011nunez}
{N{\'u}{\~n}ez}, M. 2011, Space Weather, 9, S07003, \dodoi{10.1029/2010SW000640}

\bibitem[{{N{\'u}{\~n}ez}(2015)}]{2015nunez}
---. 2015, Space Weather, 13, 807, \dodoi{10.1002/2015SW001256}

\bibitem[{Papaioannou {et~al.}(2018)Papaioannou, Anastasiadis, Kouloumvakos, Paassilta, Vainio, Valtonen, Belov, Eroshenko, Abunina, \& Abunin}]{papaioannou2018nowcasting}
Papaioannou, A., Anastasiadis, A., Kouloumvakos, A., {et~al.} 2018, Solar Physics, 293, 100

\bibitem[{Parker(1965)}]{parker1965dynamical}
Parker, E. 1965, Space Science Reviews, 4, 666

\bibitem[{Pedregosa {et~al.}(2011)Pedregosa, Varoquaux, Gramfort, Michel, Thirion, Grisel, Blondel, Prettenhofer, Weiss, Dubourg, Vanderplas, Passos, Cournapeau, Brucher, Perrot, \& Duchesnay}]{scikit-learn}
Pedregosa, F., Varoquaux, G., Gramfort, A., {et~al.} 2011, Journal of Machine Learning Research, 12, 2825

\bibitem[{Perlich {et~al.}(2003)Perlich, Provost, \& Simonoff}]{perlich2003tree}
Perlich, C., Provost, F., \& Simonoff, J. 2003, Journal of Machine Learning Research, 211

\bibitem[{Posner(2007)}]{posner2007up}
Posner, A. 2007, Space Weather, 5

\bibitem[{Reames(1999)}]{reames1999particle}
Reames, D.~V. 1999, Space Science Reviews, 90, 413

\bibitem[{Rotti {et~al.}(2022)Rotti, Aydin, Georgoulis, \& Martens}]{gsep_2022}
Rotti, S., Aydin, B., Georgoulis, M., \& Martens, P. 2022, {GSEP Dataset}, V5,  Harvard Dataverse, \dodoi{10.7910/DVN/DZYLHK}

\bibitem[{{Rotti} {et~al.}(2022){Rotti}, {Aydin}, {Georgoulis}, \& {Martens}}]{rotti2022}
{Rotti}, S., {Aydin}, B., {Georgoulis}, M.~K., \& {Martens}, P.~C. 2022, \apjs, 262, 29, \dodoi{10.3847/1538-4365/ac87ac}

\bibitem[{Rotti \& Martens(2023)}]{rotti2023analysis}
Rotti, S., \& Martens, P.~C. 2023, The Astrophysical Journal Supplement Series, 267, 40

\bibitem[{Ruiz {et~al.}(2021)Ruiz, Flynn, Large, Middlehurst, \& Bagnall}]{ruiz2021great}
Ruiz, A.~P., Flynn, M., Large, J., Middlehurst, M., \& Bagnall, A. 2021, Data Mining and Knowledge Discovery, 35, 401

\bibitem[{Sadykov {et~al.}(2021)Sadykov, Kosovichev, Kitiashvili, Oria, Nita, Illarionov, O'Keefe, Jiang, Fereira, \& Ali}]{sadykov2021prediction}
Sadykov, V., Kosovichev, A., Kitiashvili, I., {et~al.} 2021, arXiv preprint arXiv:2107.03911

\bibitem[{Safavian \& Landgrebe(1991)}]{safavian1991survey}
Safavian, S.~R., \& Landgrebe, D. 1991, IEEE transactions on systems, man, and cybernetics, 21, 660

\bibitem[{Sanner {et~al.}(1999)}]{sanner1999python}
Sanner, M.~F., {et~al.} 1999, J Mol Graph Model, 17, 57

\bibitem[{Sauer(1989)}]{sauer1989sel}
Sauer, H.~H. 1989, SEL monitoring of the earth’s energetic particle radiation environment

\bibitem[{Schaefer(1990)}]{schaefer1990critical}
Schaefer, J.~T. 1990, Weather and forecasting, 5, 570

\bibitem[{Sch{\"a}fer(2015)}]{schafer2015boss}
Sch{\"a}fer, P. 2015, Data Mining and Knowledge Discovery, 29, 1505

\bibitem[{Sch{\"a}fer \& H{\"o}gqvist(2012)}]{schafer2012sfa}
Sch{\"a}fer, P., \& H{\"o}gqvist, M. 2012, in Proceedings of the 15th international conference on extending database technology, 516--527

\bibitem[{Schapire(1990)}]{schapire1990strength}
Schapire, R.~E. 1990, Machine learning, 5, 197

\bibitem[{Schrijver \& Siscoe(2010)}]{schrijver2010heliophysics}
Schrijver, C.~J., \& Siscoe, G.~L. 2010, Heliophysics: space storms and radiation: causes and effects (Cambridge University Press)

\bibitem[{Schwadron {et~al.}(2010)Schwadron, Townsend, Kozarev, Dayeh, Cucinotta, Desai, Golightly, Hassler, Hatcher, Kim, {et~al.}}]{schwadron2010earth}
Schwadron, N.~A., Townsend, L., Kozarev, K., {et~al.} 2010, Space Weather, 8

\bibitem[{Singer {et~al.}(2001)Singer, Heckman, \& Hirman}]{singer2001space}
Singer, H., Heckman, G., \& Hirman, J. 2001, Washington DC American Geophysical Union Geophysical Monograph Series, 125, 23

\bibitem[{Smart \& Shea(1992)}]{smart1992}
Smart, D., \& Shea, M. 1992, Advances in Space Research, 12, 303

\bibitem[{Swalwell {et~al.}(2017)Swalwell, Dalla, \& Walsh}]{swalwell2017solar}
Swalwell, B., Dalla, S., \& Walsh, R.~W. 2017, Solar Physics, 292, 1

\bibitem[{Van Der~Walt {et~al.}(2011)Van Der~Walt, Colbert, \& Varoquaux}]{van2011numpy}
Van Der~Walt, S., Colbert, S.~C., \& Varoquaux, G. 2011, Computing in science \& engineering, 13, 22

\bibitem[{{Van Hollebeke} {et~al.}(1975){Van Hollebeke}, {Ma Sung}, \& {McDonald}}]{1975van}
{Van Hollebeke}, M.~A.~I., {Ma Sung}, L.~S., \& {McDonald}, F.~B. 1975, \solphys, 41, 189, \dodoi{10.1007/BF00152967}

\bibitem[{Whitman {et~al.}(2022)Whitman, Egeland, Richardson, Allison, Quinn, Barzilla, Kitiashvili, Sadykov, Bain, Dierckxsens, Mays, Tadesse, Lee, Semones, Luhmann, Núñez, White, Kahler, Ling, Smart, Shea, Tenishev, Boubrahimi, Aydin, Martens, Angryk, Marsh, Dalla, Crosby, Schwadron, Kozarev, Gorby, Young, Laurenza, Cliver, Alberti, Stumpo, Benella, Papaioannou, Anastasiadis, Sandberg, Georgoulis, Ji, Kempton, Pandey, Li, Hu, Zank, Lavasa, Giannopoulos, Falconer, Kadadi, Fernandes, Dayeh, Muñoz-Jaramillo, Chatterjee, Moreland, Sokolov, Roussev, Taktakishvili, Effenberger, Gombosi, Huang, Zhao, Wijsen, Aran, Poedts, Kouloumvakos, Paassilta, Vainio, Belov, Eroshenko, Abunina, Abunin, Balch, Malandraki, Karavolos, Heber, Labrenz, Kühl, Kosovichev, Oria, Nita, Illarionov, O’Keefe, Jiang, Fereira, Ali, Paouris, Aminalragia-Giamini, Jiggens, Jin, Lee, Palmerio, Bruno, Kasapis, Wang, Chen, Sanahuja, Lario, Jacobs, Strauss, Steyn, {van den Berg}, Swalwell, Waterfall, Nedal, Miteva, Dechev, Zucca, Engell,
  Maze, Farmer, Kerber, Barnett, Loomis, Grey, Thompson, Linker, Caplan, Downs, Török, Lionello, Titov, Zhang, \& Hosseinzadeh}]{WHITMAN2022}
Whitman, K., Egeland, R., Richardson, I.~G., {et~al.} 2022, Advances in Space Research, \dodoi{https://doi.org/10.1016/j.asr.2022.08.006}

\bibitem[{Wilks(1990)}]{wilks1990combination}
Wilks, D.~S. 1990, Weather and forecasting, 5, 640

\bibitem[{{Winter} \& {Ledbetter}(2015)}]{winter2015ApJ}
{Winter}, L.~M., \& {Ledbetter}, K. 2015, \apj, 809, 105, \dodoi{10.1088/0004-637X/809/1/105}

\bibitem[{Woodcock(1976)}]{woodcock_evaluation_1976}
Woodcock, F. 1976, Monthly Weather Review, 104, 1209 , \dodoi{10.1175/1520-0493(1976)104<1209:TEOYFF>2.0.CO;2}

\bibitem[{Youden(1950)}]{youden1950index}
Youden, W.~J. 1950, Cancer, 3, 32

\bibitem[{Zhang {et~al.}(2010)Zhang, Jin, \& Zhou}]{zhang2010understanding}
Zhang, Y., Jin, R., \& Zhou, Z.-H. 2010, International journal of machine learning and cybernetics, 1, 43

\end{thebibliography}
\bibliographystyle{aasjournal}


\end{document}